\documentclass[12pt]{article}
\usepackage[margin=1.9cm]{geometry}
\usepackage{epsfig}
\usepackage{cite}

\newcommand{\mysection}{\setcounter{equation}{0}\section}

\def\beq{\begin{equation}}
\def\eeq{\end{equation}}
\def\beqa{\begin{eqnarray}}
\def\eeqa{\end{eqnarray}}
 
\begin{document}

\begin{center}
{\Large \bf Higher-order corrections for $tW$ production \\ at high-energy hadron colliders}
\end{center}
\vspace{2mm}
\begin{center}
{\large Nikolaos Kidonakis and Nodoka Yamanaka}\\
\vspace{2mm}
{\it Department of Physics, Kennesaw State University,\\
Kennesaw, GA 30144, USA}
\end{center}
 
\begin{abstract}
We discuss cross sections for $tW$ production in proton-proton collisions at the LHC and at higher-energy colliders with energies of up to 100 TeV. We find that, remarkably, the soft-gluon corrections are numerically dominant even at very high collider energies. We present results with soft-gluon corrections at approximate NNLO and approximate N$^3$LO matched to complete NLO results. These higher-order corrections are large and need to be included for better theoretical accuracy and smaller scale dependence. Total cross sections as well as top-quark and $W$-boson transverse-momentum and rapidity distributions are presented using various recent sets of parton distribution functions.

\end{abstract}
 
\mysection{Introduction}

The associated production of a top quark with a $W$ boson is an important process that has been studied extensively at the Large Hadron Collider (LHC) and is expected to play a major role in future colliders. The underlying partonic process for $tW$ production at hadron colliders is $bg \to tW^-$. This process is critically dependent on the value of the CKM matrix element $V_{tb}$, and it could be affected by new physics beyond the Standard Model.

Results for $tW$ production and decays at leading order (LO) were presented some time ago in Refs. \cite{LY,HBB,SM,BBD}, and some additional corrections appeared in \cite{TT,BB}. Next-to-leading-order (NLO) calculations of the QCD corrections for this process were derived in Ref. \cite{Zhu}. Further NLO studies of $tW$ production including the decays of the top quark and the $W$ boson were done in Ref. \cite{JCFT}. The top-quark transverse-momentum ($p_T$) distributions for this process at NLO matched with parton showers were presented in Ref. \cite{Re}. Some updated results for the NLO total cross section appeared in Ref. \cite{HATHOR}. In all these studies, it has been shown that the NLO corrections are very significant and, thus, their inclusion is required to make meaningful theoretical predictions for comparison to collider data. 

Given the large size of the NLO corrections, it is important to consider even higher-order corrections. Additional corrections beyond NLO from soft-gluon emission at various logarithmic accuracies were presented in Refs. \cite{NKsingletopTev,NKsingletopLHC,NKtWH,NKpT,NKtW}. Next-to-leading-logarithm (NLL) resummation for single-top production processes, including $tW$ production, was first derived in Ref. \cite{NKsingletopTev}. NLL resummation requires one-loop calculations of the corresponding soft anomalous dimensions which were performed in \cite{NKsingletopTev}. Fixed-order expansions of the resummed cross section and applications to Tevatron energies were also provided in Ref. \cite{NKsingletopTev}. A study of $tW$ production and other single-top processes at LHC energies followed in Ref. \cite{NKsingletopLHC}.

A two-loop calculation of the soft anomalous dimension for $tW$ production and the derivation of next-to-next-to-leading-logarithm (NNLL) resummation was presented in Ref. \cite{NKtWH}. Fixed-order expansions at approximate next-to-next-to-leading order (aNNLO) and a study at LHC energies was also given in \cite{NKtWH}. The top-quark $p_T$ distributions in single-top production processes, including $tW$ production, were studied at aNNLO at LHC energies in Ref. \cite{NKpT}.

Approximate next-to-next-to-next-to-leading order (aN$^3$LO) results for $tW$ production from NNLL resummation at LHC (and Tevatron) energies were given in Ref. \cite{NKtW}. Results were given for the total cross section, the top-quark $p_T$ distributions, and the top-quark rapidity distributions. The theoretical predictions for the cross section were compared in Ref. \cite{NKtW} - as well as in the review paper in Ref. \cite{NKtoprev} - with data from the LHC \cite{ATLAS7,CMS7,CMS8,ATLAS8,LHC8,ATLAS13,CMS13}, finding very good agreement. More recently, three-loop calculations for the soft anomalous dimensions in single-top production, including $tW$ production, were presented in Ref. \cite{NK3loop} (see also a recent review paper on soft anomalous dimensions for QCD processes \cite{NKuni}). 

In this paper we explore $tW$ production at collider energies of up to 100 TeV using the latest theoretical results. We calculate the soft-gluon corrections through third order and match to complete NLO calculations to produce aNNLO and aN$^3$LO theoretical predictions. We find that, remarkably, the soft-gluon corrections are numerically dominant throughout the energy range that we study, even at very high collider energies. Thus, soft-gluon resummation is much more useful and applicable than just for calculations near threshold. We also update predictions for LHC energies, using the latest theoretical input, and compare with recent data from the LHC. In Section 2 we briefly review the soft-gluon formalism for $tW$ production. In Section 3 we provide results for total cross sections over a wide range of collision energies, and in Section 4 we present top-quark and $W$-boson differential distributions in transverse momentum and rapidity. We conclude in Section~5.

\mysection{Soft-gluon resummation for $tW$ production}

The formalism for soft-gluon resummation will be briefly described in this section (see Refs. \cite{NKsingletopTev,NKtWH,NKtW,NKtoprev,NK3loop,NKuni,GS87,CT89,NKGS1,NKGS2,KOS,LOS,ADS,HRSV,MFNK}). We study the process $b(p_1)\, + \, g\, (p_2) \to t(p_3)\, + W^-(p_4)$ and define the usual kinematical variables $s=(p_1+p_2)^2$, $t=(p_1-p_3)^2$, $u=(p_2-p_3)^2$. We denote the top-quark mass by $m_t$ and the $W$-boson mass by $m_W$. We also define the partonic threshold variable $s_4=s+t+u-m_t^2-m_W^2$ which measures the energy in the soft-gluon radiation and which vanishes at threshold. 

We write the differential cross section, $d\sigma_{bg \to tW}$, as
\beq
d\sigma_{bg \to tW}= \int dx_1 \, dx_2 \,  \phi(x_1, \mu_F) \, \phi(x_2, \mu_F) \, 
d{\hat \sigma}_{bg \to tW}(s_4, \alpha_s, \mu_F,\mu_R) \, ,
\label{facphi}
\eeq
where the $\phi$ are parton distribution functions (pdf) for the bottom quark and the gluon (we omit subscripts for simplicity), and ${\hat \sigma}_{bg \to tW}$ is the hard-scattering partonic cross section. The cross section depends on the strong coupling, $\alpha_s$, the factorization scale, $\mu_F$, and the renormalization scale, $\mu_R$.

We define Laplace transforms, with variable $N$, of the partonic cross section as $d{\tilde{\hat\sigma}}_{bg \to tW}(N)=\int_0^s (ds_4/s) \,  e^{-N s_4/s} \, d{\hat\sigma}_{bg \to tW}(s_4)$. Under the Laplace transform, the logarithms of $s_4$ in the perturbative expansion turn into logarithms of $N$. We also define transforms of the parton distributions through the relation ${\tilde \phi}(N)=\int_0^1 e^{-N(1-x)} \phi(x) \, dx$. We then find the transform-space expression
\beq
d{\tilde\sigma}_{bg \to tW}(N)= {\tilde \phi}(N_1) \, {\tilde \phi}(N_2) \, 
d{\tilde{\hat \sigma}}_{bg \to tW}(N) \, ,
\label{facphiN}
\eeq
where $N_1=N(m_W^2-u)/s$ and $N_2=N(m_W^2-t)/s$.

The cross section for $tW$ production can be rewritten under Laplace transforms in refactorized form as a product of separate functions: a short-distance, infrared safe, hard function, $H_{bg \to tW}$; a soft function, $S_{bg \to tW}$, which describes the emission of noncollinear soft gluons in the process; and functions $\psi$ that describe collinear emission from the incoming partons \cite{NKtW,NKtoprev,NKuni,NKGS1,NKGS2,KOS,LOS,ADS,HRSV,MFNK}. We have
\beqa
d{\tilde \sigma}_{bg \to tW}(N)&=& 
{\tilde \psi}(N_1) \, {\tilde \psi}(N_2) \,
H_{bg \to tW}\left(\alpha_s(\mu_R)\right) \, 
{\tilde S}_{bg \to tW}\left(\frac{\sqrt{s}}{N \mu_F} \right) \, .
\label{refac}
\eeqa

By comparing Eq. (\ref{facphiN}) with Eq. (\ref{refac}), we find a new expression for the hard-scattering partonic cross section in $N$ space:
\beq
d{\tilde{\hat \sigma}}_{bg \to tW}(N)=
\frac{{\tilde \psi}(N_1) \, {\tilde \psi}(N_2)} {{\tilde \phi}(N_1) \, 
{\tilde \phi}(N_2)} \, H_{bg \to tW}\left(\alpha_s(\mu_R)\right) \, 
{\tilde S}_{bg \to tW}\left(\frac{\sqrt{s}}{N \mu_F} \right) \, .
\label{sigN}
\eeq

The soft function ${\tilde S}_{bg \to tW}$ observes the renormalization-group equation
\beq
\left(\mu_R \frac{\partial}{\partial \mu_R}
+\beta(g_s)\frac{\partial}{\partial g_s}\right) {\tilde S}_{bg \to tW}
=-2\, {\tilde S}_{bg \to tW} \; \Gamma_{\! S \, bg \to tW}
\eeq
where $g_s^2=4\pi\alpha_s$ and $\beta$ is the QCD beta function. The soft anomalous dimension, $\Gamma_{\! S \, bg \to tW}$, controls the evolution of the soft function, and it is calculated from the coefficients of the ultraviolet poles of the relevant eikonal diagrams \cite{NKsingletopTev,NKtWH,NK3loop}.

The resummed differential cross section in transform space is derived from the renormalization-group evolution of the soft function and the ${\tilde \psi}/{\tilde \phi}$ ratios in Eq. (\ref{sigN}). We have \cite{NKGS1,NKGS2,LOS,NKsingletopTev,NKtWH,NKtoprev,NK3loop,NKuni}
\beqa
d{\tilde{\hat \sigma}}^{\rm resum}_{bg \to tW}(N) &=&   
\exp\left[\sum_{i=b,g} E_i(N_i)\right] \; 
H_{bg \to tW}\left(\alpha_s(\sqrt{s})\right) 
\nonumber\\ && \times
{\tilde S}_{bg \to tW} \left(\alpha_s(\frac{\sqrt{s}}{\tilde N})\right) \; 
\exp \left[2\int_{\sqrt{s}}^{{\sqrt{s}}/{\tilde N}} 
\frac{d\mu}{\mu} \Gamma_{\! S \, bg \to tW}
\left(\alpha_s(\mu)\right)\right] \, . 
\label{restW}
\eeqa

The first exponent in the above expression resums soft and collinear corrections from the incoming partons \cite{GS87,CT89} (see e.g. Ref. \cite{NKuni} for explicit expressions). 
The soft anomalous dimension has a perturbative expansion $\Gamma_{\! S \, bg \to tW}=\sum_{n=1}^{\infty} (\alpha_s/\pi)^n \, \Gamma_{\! S \, bg \to tW}^{(n)}$, and it has been calculated to three loops.

The soft anomalous dimension for $bg \to tW^-$ is given at one loop by  \cite{NKsingletopTev,NKtWH}
\beq
\Gamma_{\! S \, bg \to tW}^{(1)}=C_F \left[\ln\left(\frac{m_t^2-t}{m_t\sqrt{s}}\right)
-\frac{1}{2}\right] +\frac{C_A}{2} \ln\left(\frac{u-m_t^2}{t-m_t^2}\right) \, ,
\eeq
where $C_F=(N_c^2-1)/(2 N_c)$ and $C_A=N_c$, with $N_c=3$ the number of colors in QCD.

The two-loop soft anomalous dimension is given by \cite{NKtWH}
\beq
\Gamma_{\! S \, bg \to tW}^{(2)}=K_2 \, \Gamma_{\!\! S \, bg \to tW}^{(1)}
+\frac{1}{4}C_F C_A (1-\zeta_3) \, ,
\eeq
with 
\beq
K_2=C_A \left(\frac{67}{36}-\frac{\zeta_2}{2}\right)-\frac{5}{18} n_f \, ,
\eeq
$\zeta_2=\pi^2/6$, $\zeta_3=1.202056903\cdots$, and $n_f$ the number of light-quark flavors.

The three-loop soft anomalous dimension is given by \cite{NK3loop}
\beq
\Gamma_{\! S \, bg \to tW}^{(3)}=K_3 \, \Gamma_{\!\! S \, bg \to tW}^{(1)}+\frac{1}{2} K_2 C_F C_A (1-\zeta_3)+C_F C_A^2\left(-\frac{1}{4}+\frac{3}{8}\zeta_2-\frac{\zeta_3}{8}-\frac{3}{8}\zeta_2 \zeta_3+\frac{9}{16} \zeta_5\right) \, ,
\eeq
where 
\beq
K_3=C_A^2\left(\frac{245}{96}-\frac{67}{36}\zeta_2
+\frac{11}{24}\zeta_3+\frac{11}{8}\zeta_4\right)
+C_F n_f\left(-\frac{55}{96}+\frac{\zeta_3}{2}\right)
+C_A n_f \left(-\frac{209}{432}+\frac{5}{18}\zeta_2
-\frac{7}{12}\zeta_3\right)-\frac{n_f^2}{108} 
\eeq
and $\zeta_4=\pi^4/90$, $\zeta_5=1.036927755\cdots$.

The $N$-space resummed cross section in Eq. (\ref{restW}) can be expanded to any fixed order and then inverted back to momentum space. The soft-gluon corrections appear in the perturbative expansion for the physical cross section in terms of plus distributions that involve logarithms of $s_4$, i.e., $[(\ln^k(s_4/m_t^2))/s_4]_+$, where $k$ takes values from 0 to $2n-1$ for the $n$th order corrections in the strong coupling, $\alpha_s$. These plus distributions are defined by their integrals with smooth functions $f$ as 
\beqa
\int_0^{s_4^{max}} ds_4 \, \left[\frac{\ln^k(s_4/m_t^2)}
{s_4}\right]_{+} f(s_4) &=&
\int_0^{s_4^{max}} ds_4 \frac{\ln^k(s_4/m_t^2)}{s_4} [f(s_4) - f(0)]
\nonumber \\ &&
{}+\frac{1}{k+1} \ln^{k+1}\left(\frac{s_4^{max}}{m_t^2}\right) f(0) \, .
\label{plus}
\eeqa

The soft-gluon corrections at NLO are   
\beqa
\frac{d\hat{\sigma}_{bg \to tW}^{(1)}}{dt \, du}&=&F_{bg \to tW}^{\rm LO} 
\frac{\alpha_s(\mu_R)}{\pi} \;  
\left\{2(C_F+C_A) \; \left[\frac{\ln(s_4/m_t^2)}{s_4}\right]_+ \right.
\nonumber \\ && \hspace{-15mm} 
{}+\left[-2 C_F \, \ln\left(\frac{m_W^2-u}{m_t^2}\right)
-2 C_A \, \ln\left(\frac{m_W^2-t}{m_t^2}\right)
-(C_F+C_A) \ln\left(\frac{\mu_F^2}{s}\right)
+2 \, \Gamma_{\! S \, bg \to tW}^{(1)}\right] \; 
\left[\frac{1}{s_4}\right]_+ 
\nonumber \\ && \hspace{-15mm} \left. 
{}+\left[\left(C_F\, \ln\left(\frac{m_W^2-u}{m_t^2}\right) 
+C_A \, \ln\left(\frac{m_W^2-t}{m_t^2}\right)
-\frac{3}{4}C_F-\frac{\beta_0}{4}\right)\ln\left(\frac{\mu_F^2}{m_t^2}\right) 
+\frac{\beta_0}{4} \ln\left(\frac{\mu_R^2}{m_t^2}\right) \right] \delta(s_4)
\right\} \, ,
\nonumber \\
\label{aNLO}
\eeqa
where $\beta_0=(11C_A-2n_f)/3$ and 
\beqa
F_{bg \to tW}^{\rm LO}&=& 
\frac{\pi \, V_{tb}^2 \, \alpha_s \, \alpha}{12 m_W^2 \sin^2\theta_W s^2}
\left\{-\frac{(2 m_W^2+m_t^2)}{2 (u-m_t^2)^2} 
\left[(u-m_W^2)(s+3m_t^2-m_W^2)+(t-m_t^2) (m_t^2- m_W^2)\right] \right.
\nonumber \\ && \hspace{-10mm}
{}+\frac{1}{s(u-m_t^2)}\left[2(t-m_t^2)(m_t^2-m_W^2)m_W^2+(u-m_W^2) (t+u-2m_t^2) m_t^2+s m_t^2 (2 m_W^2+m_t^2)\right]
\nonumber \\ && \hspace{-10mm} \left. 
{}-\frac{(u-m_t^2)}{2s} \left(2 m_W^2+m_t^2\right) \right\} \, ,
\eeqa
with $\alpha=e^2/(4\pi)$, $\theta_W$ the weak mixing angle, and $V_{tb}$ a CKM matrix element. For explicit analytical expressions at higher orders see Refs. \cite{NKtW,NKtoprev,NKuni}. 

\mysection{Total cross sections for $tW$ production}

In this section we present numerical results for $tW$ production in $pp$ collisions. We present total cross sections for a large range of collider energies.

\begin{figure}[htbp]
\begin{center}
\includegraphics[width=90mm]{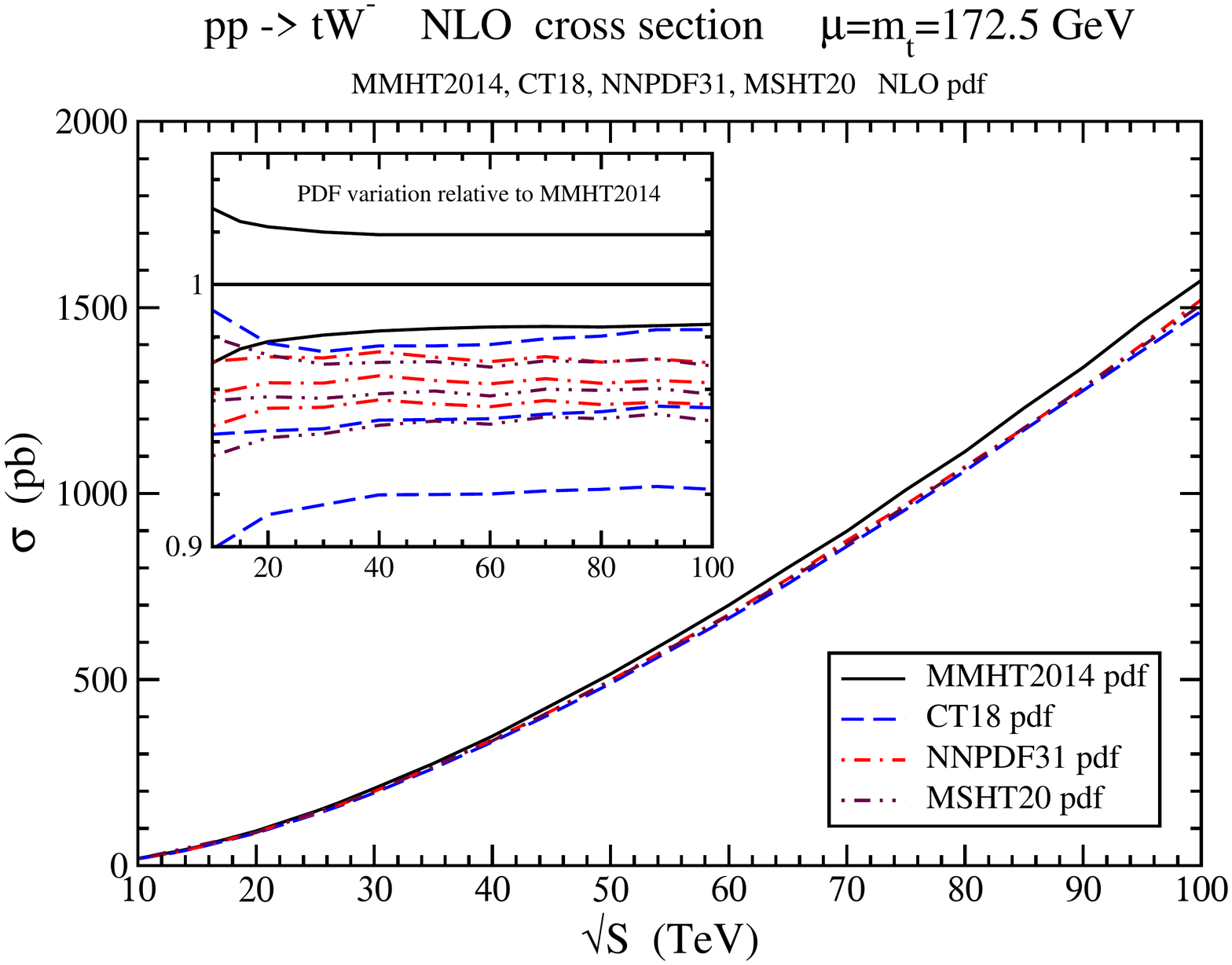}
\hspace{-7mm}
\includegraphics[width=90mm]{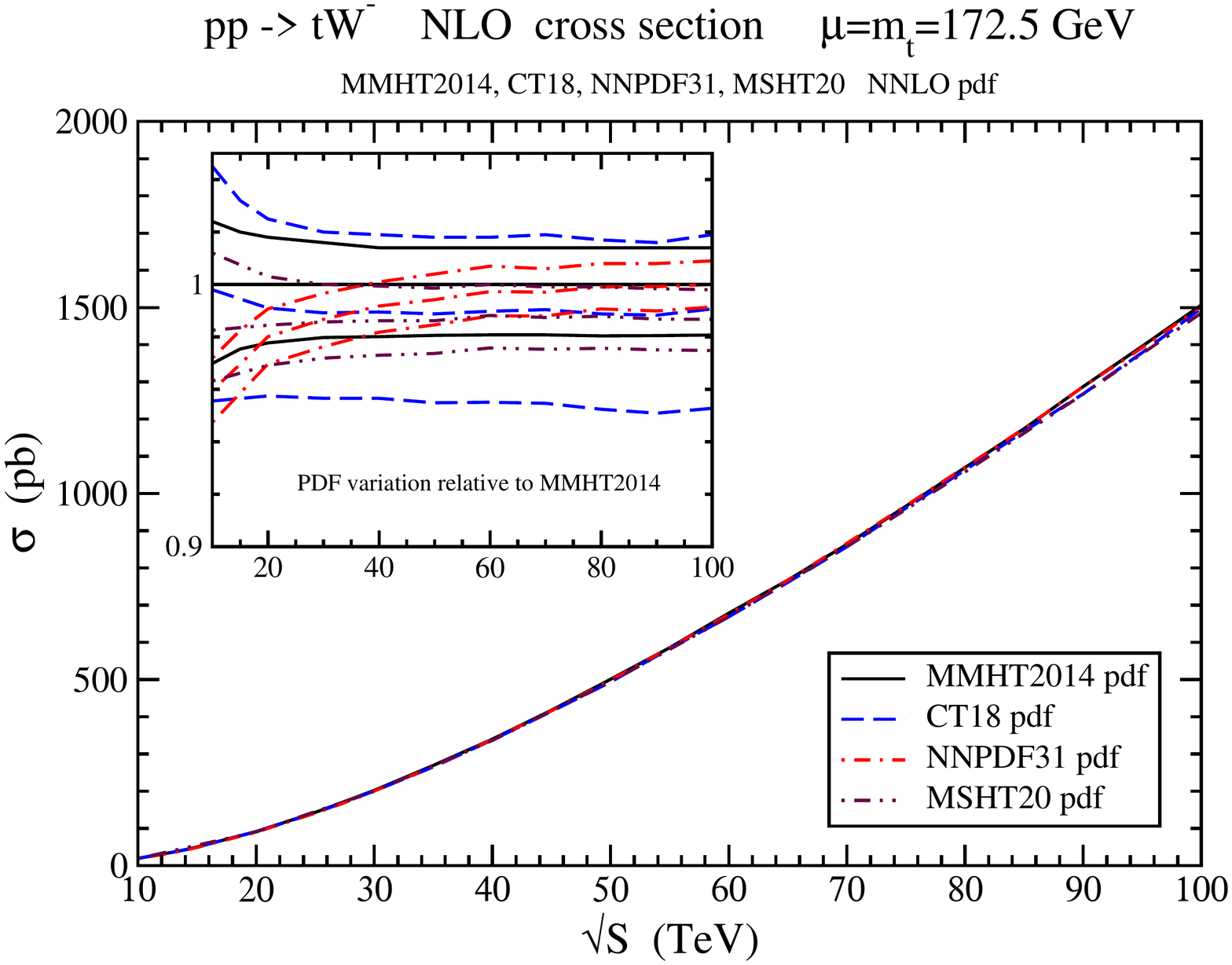}
\caption{The NLO cross section for $tW^-$ production versus collider energy using MMHT2014, CT18, NNPDF31, and MSHT20 NLO pdf (left plot) and NNLO pdf (right plot) with $\mu=m_t=172.5$ GeV. The inset plots show the ratios of the NLO cross sections with pdf uncertainties (upper, central, lower) for the four pdf sets to the central MMHT2014 cross section.}
\label{tWNLOpdfplot}
\end{center}
\end{figure}

In Fig. \ref{tWNLOpdfplot} we present the NLO cross section for $tW^-$ production in proton-proton collisions (the result for ${\bar t} W^+$ is the same) as a function of collision energy, $\sqrt S$, up to 100 TeV. We choose a scale $\mu$ equal to the top-quark mass, which we set as $m_t=172.5$ GeV. We have used {\small \sc MadGraph5\_aMC@NLO} \cite{MadGraph} for the complete NLO cross sections. Results are given for a number of NLO pdf sets (left plot) and NNLO pdf sets (right plot), including MMHT2014 \cite{MMHT},  NNPDF3.1 \cite{NNPDF}, CT18 \cite{CT18}, and MSHT20 \cite{MSHT}. The results with the different pdf sets are relatively close to each other. To better illustrate the difference between those pdf sets and also to show the pdf uncertainties, we display in the inset plots the ratios of the cross sections with various pdf to the cross section with the central MMHT2014 pdf (which was the main set used in Refs. \cite{NKtW,NKtoprev}). 

From the plot on the left in Fig. \ref{tWNLOpdfplot}, we observe that there are nonegligible differences in both the central results and the pdf uncertainties among the NLO pdf sets, and that the relative differences also depend on the energy. The plot on the right in Fig. \ref{tWNLOpdfplot} shows that with NNLO pdf sets the situation is much better than in the previous case. The cross sections are practically on top of each other, as can also be more clearly seen in the inset plot. Since our final goal is to obtain expressions at approximate NNLO and beyond, and thus use NNLO pdf sets, this makes the choice of pdf sets much less significant.

\begin{figure}[htbp]
\begin{center}
\includegraphics[width=90mm]{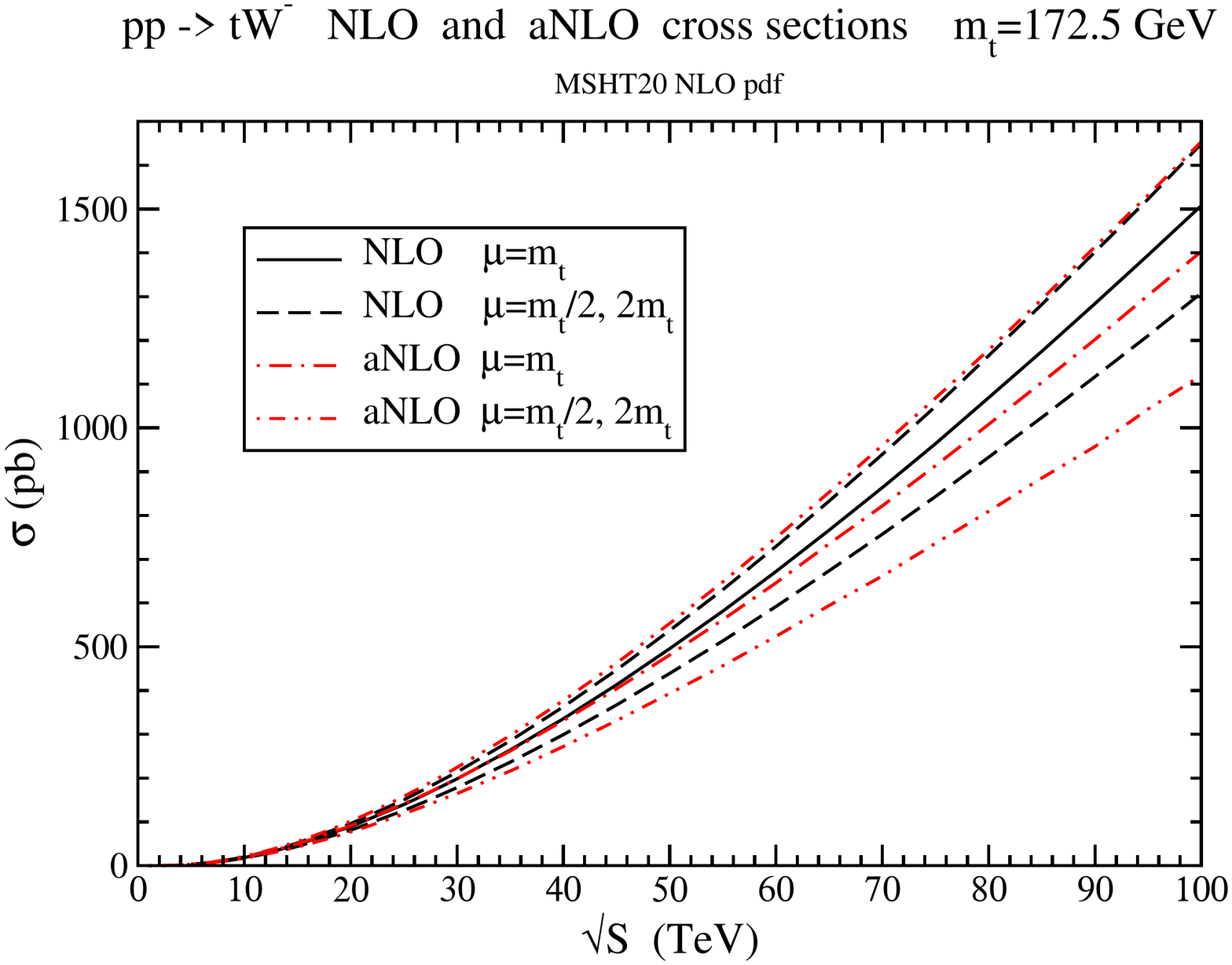}
\hspace{-7mm}
\includegraphics[width=90mm]{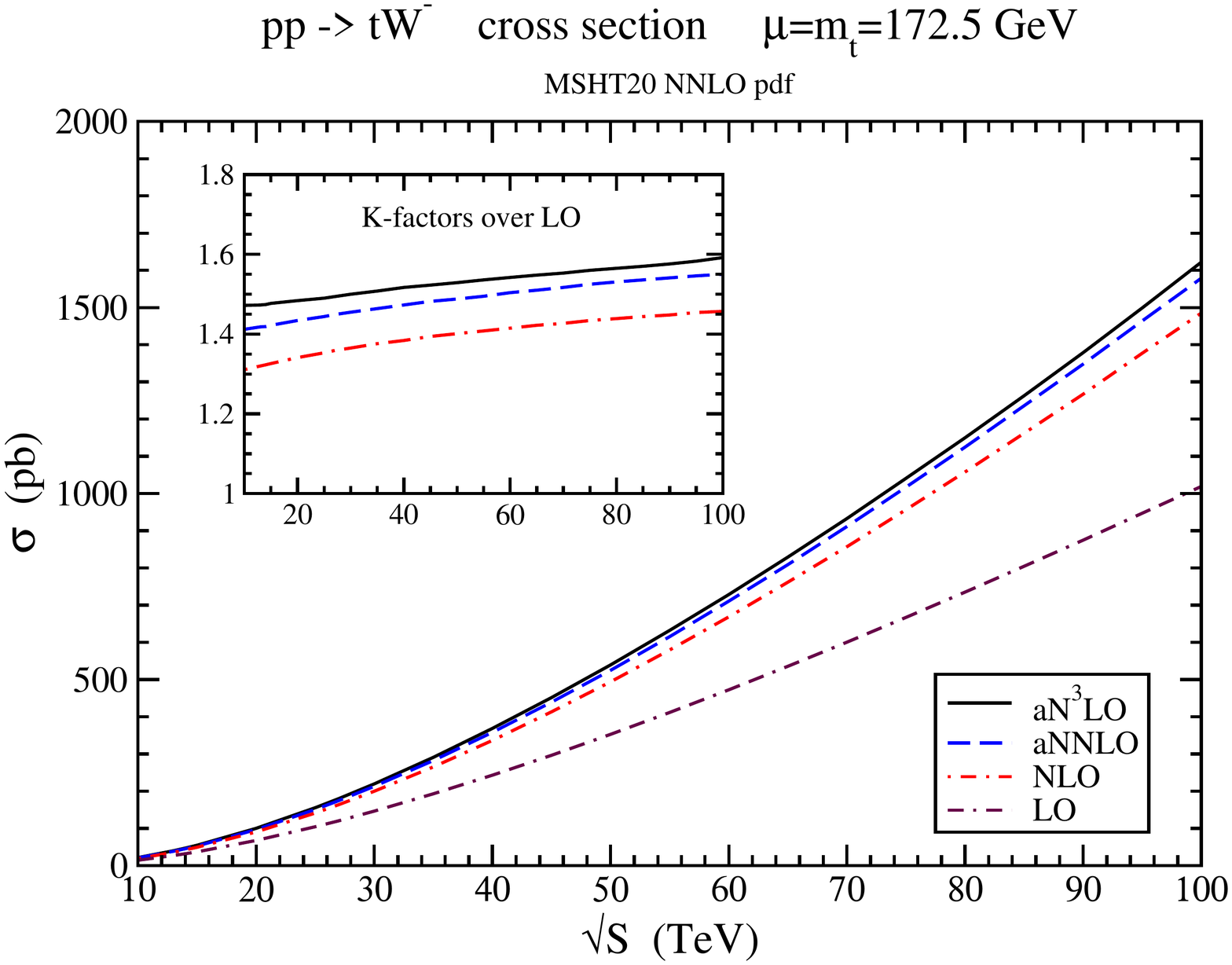}
\caption{The cross section for $tW^-$ production with $m_t=172.5$ GeV versus collider energy: (left) The NLO and aNLO cross sections using MSHT20 NLO pdf with three scale choices: $\mu=m_t/2$, $m_t$, and $2 m_t$; (right) The LO, NLO, aNNLO, and aN$^3$LO cross sections using MSHT20 NNLO pdf with $\mu=m_t$. The inset plot shows the $K$-factors over LO.}
\label{tWplot}
\end{center}
\end{figure}

We next investigate the quality and relevance of the soft-gluon approximation by comparing the complete NLO cross section with the approximate NLO (aNLO) cross section; the latter is the sum of the LO cross section and the NLO soft-gluon corrections of Eq. (\ref{aNLO}). The plot on the left in Fig. \ref{tWplot} shows the NLO and the aNLO cross sections, again as functions of $\sqrt{S}$. Results are shown using MSHT20 NLO pdf with central scale $\mu=m_t$ as well as with scales $\mu=m_t/2$ and 2$m_t$. We observe that the NLO and aNLO results are very close to each other throughout the energy range. Even at the highest collider energy of 100 TeV, the difference between the NLO and aNLO results remains small. The NLO scale variation band lies entirely within the aNLO band, with the upper bound being practically indistinguishable between NLO and aNLO even at 100 TeV.

We note that it has long been established that soft-gluon corrections are numerically dominant in a variety of top-quark and associated processes at Tevatron and LHC energies (see, for example, the discussions in Ref. \cite{NKtoprev}). However, we now establish that this numerical dominance extends considerably - at least for some processes - to much higher energies, including the energies envisioned for all foreseeable colliders for the next several decades. This is important since it provides confidence in higher-order predictions over a much larger energy region. 

In the plot on the right in Fig. \ref{tWplot} we show the LO, NLO, aNNLO, and aN$^3$LO cross sections for $tW^-$ production with $\mu=m_t$ over an energy range up to 100 TeV. MSHT20 NNLO pdf are used for all orders in the plot in order to show the growth of the perturbative series. The aNNLO cross section is derived by adding the second-order soft-gluon corrections to the complete NLO result. The aN$^3$LO cross section is derived by further adding the third-order soft-gluon corrections. The inset plot shows the $K$-factors (i.e., ratios of cross sections) relative to the LO cross section, showing that the NLO corrections are large and that the further aNNLO and aN$^3$LO corrections are also quite significant.

\begin{figure}[htbp]
\begin{center}
\includegraphics[width=90mm]{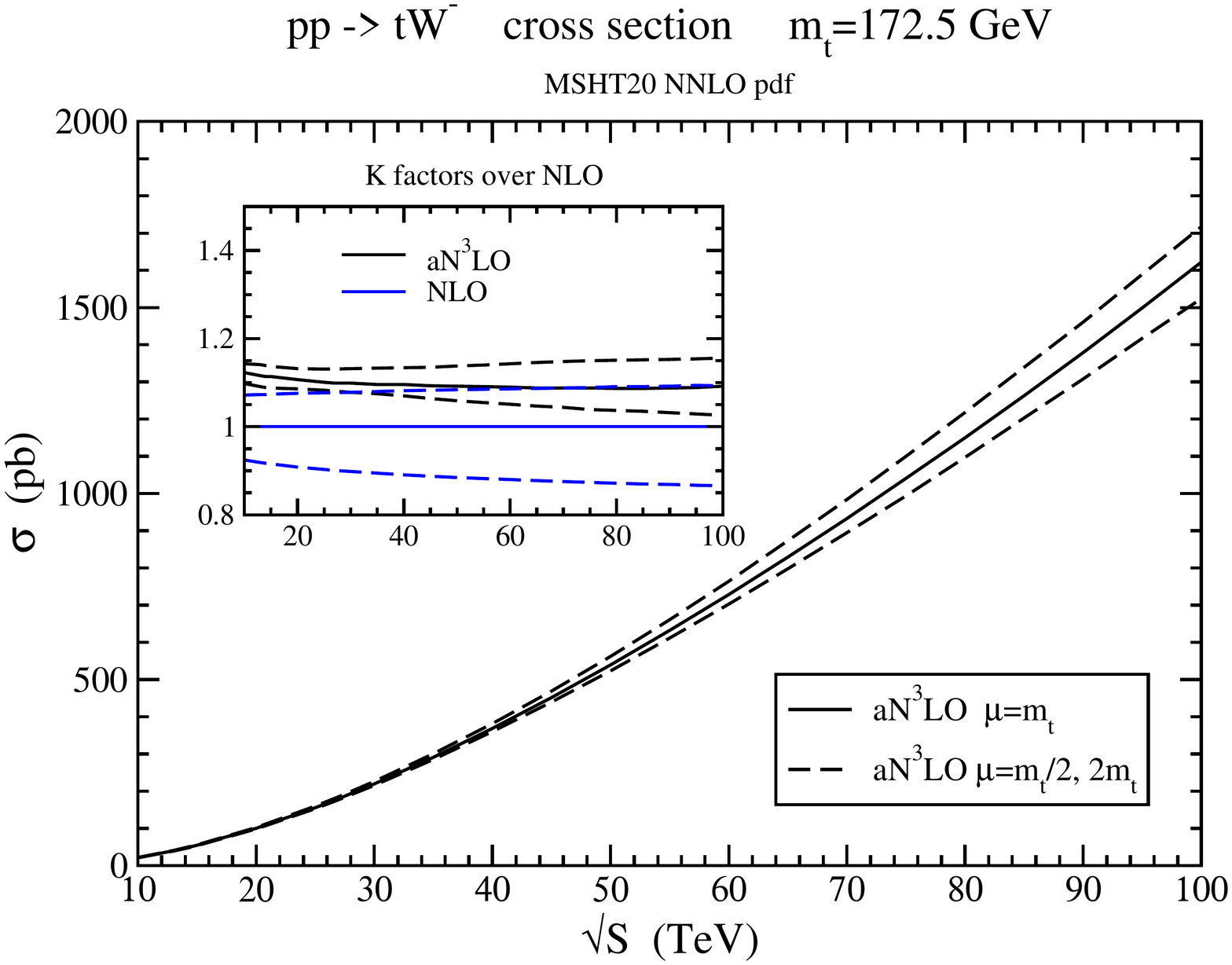}
\hspace{-7mm}
\includegraphics[width=90mm]{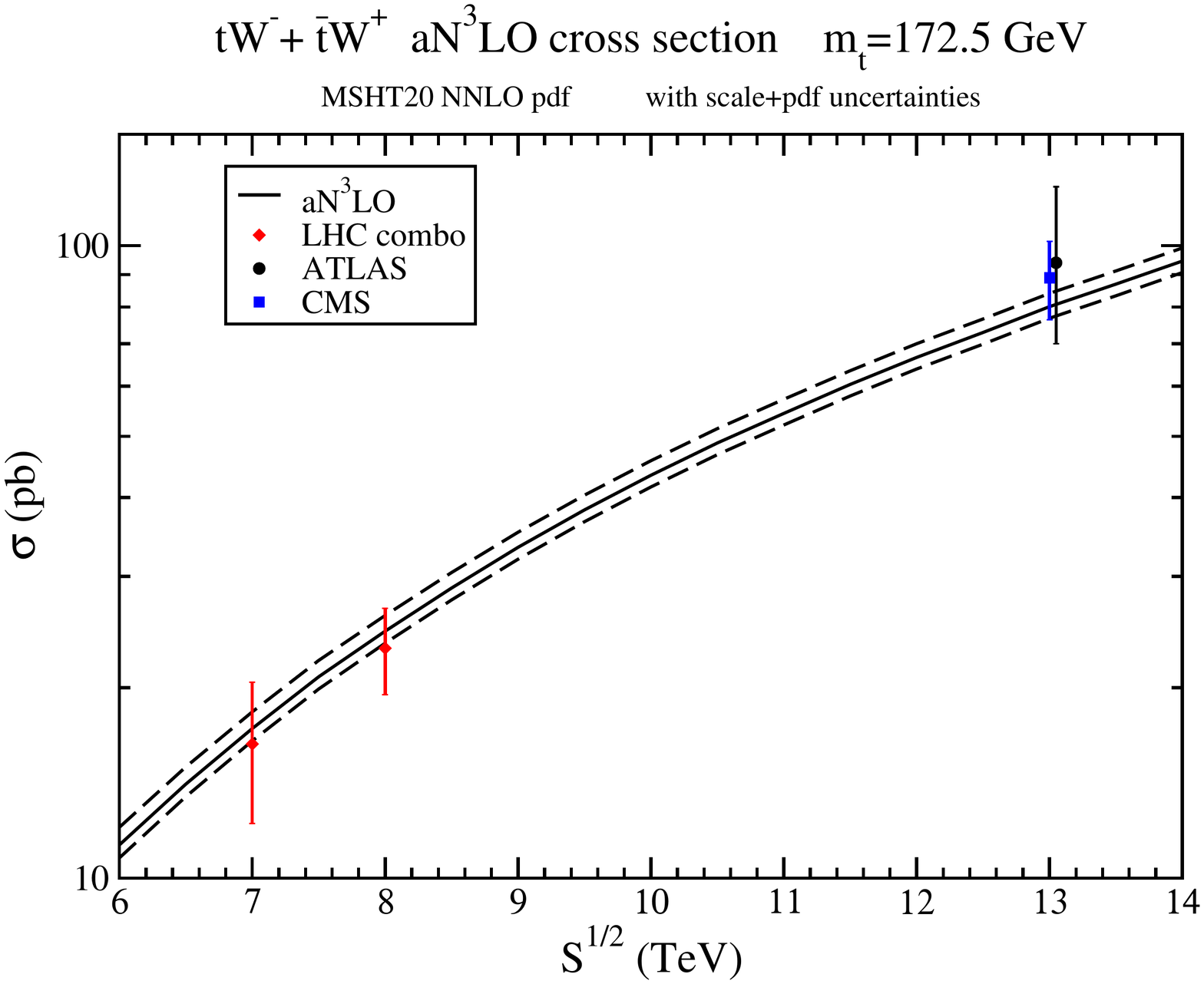}
\caption{(left) The aN$^3$LO cross section for $tW^-$ production versus collider energy using MSHT20 NNLO pdf and $m_t=172.5$ GeV with three scale choices: $\mu=m_t/2$, $m_t$, and $2 m_t$; the inset plot displays $K$-factors for the NLO and aN$^3$LO cross sections with scale variation over the central ($\mu=m_t$) NLO result. (right) Results for the  $tW^-+{\bar t}W^+$ cross section at aN$^3$LO with scale and pdf uncertainties compared with data from the LHC \cite{ATLASCMS7and8,ATLAS13,CMS13n}.}
\label{tWscalelhcplot}
\end{center}
\end{figure}

The plot on the left in Fig. \ref{tWscalelhcplot} displays results for the aN$^3$LO cross section as a function of collider energy with three different choices of factorization/renormalization scale:  $\mu=m_t/2$, $m_t$, and $2 m_t$. To better show the dependence on the scale, the inset plot shows the $K$-factors for the NLO and aN$^3$LO cross sections relative to the central NLO result. It can be seen that the scale dependence is significantly reduced at aN$^3$LO relative to NLO. This is of course important for providing more robust theoretical predictions in making comparisons with experimental data from the LHC and future colliders. We also note that the relative scale dependence depends on the collision energy, and it increases at larger energies.

The plot on the right in Fig. \ref{tWscalelhcplot} displays aN$^3$LO results for the sum of the $tW^-$ and ${\bar t}W^+$ cross sections (which is double that for $tW^-$ alone) at LHC energies as well as the relevant data from LHC combinations at 7 and 8 TeV \cite{ATLASCMS7and8} and from ATLAS \cite{ATLAS13} and CMS \cite{CMS13n} at 13 TeV. The central aN$^3$LO cross section is displayed as well as upper and lower results that include the combined scale and pdf uncertainties. We observe very good agreement between the aN$^3$LO theoretical predictions and the data from the LHC. We also note that the theoretical uncertainty is smaller than the uncertainty of the recent LHC data.

\begin{table}[htb]
\begin{center}
\begin{tabular}{|c|c|c|c|} \hline
\multicolumn{4}{|c|}{aN$^3$LO $tW^-+{\bar t}W^+$ cross sections} \\ \hline
LHC 7 TeV & LHC 8 TeV & LHC 13 TeV & LHC 14 TeV \\ \hline
$17.2 {}^{+0.4}_{-0.3} {}^{+0.7}_{-0.4}$ pb & 
$24.6 {}^{+0.6}_{-0.5} {}^{+0.9}_{-0.6}$ pb &
$79.5 {}^{+1.9}_{-1.8} {}^{+2.0}_{-1.4}$ pb &
$94.0 {}^{+2.2}_{-2.1} {}^{+2.2}_{-1.6}$ pb \\ \hline
\end{tabular}
\caption[]{The aN$^3$LO $tW^-+{\bar t}W^+$ cross sections (with scale and pdf uncertainties) in $pp$ collisions at the LHC with $\sqrt{S}=7$, 8, 13, and 14 TeV, with $m_t=172.5$ GeV and MSHT20 NNLO pdf.}
\label{table1}
\end{center}
\end{table}

Next, we provide some specific numbers for the aN$^3$LO cross sections and their uncertainties for current and future LHC energies as well as for a couple of possible energies at future colliders. Table~1 shows the aN$^3$LO cross sections for the sum of the $tW^-$ and ${\bar t} W^+$ cross sections at LHC energies. The central value is with $\mu=m_t$, the first uncertainty is from scale variation over $m_t/2 \le \mu \le 2m_t$, and the second uncertainty is from the MSHT20 NNLO pdf as provided by that set. In addition, we calculate the corresponding  $tW^-+{\bar t} W^+$ cross section at 50 TeV energy, and find $1.08^{+0.05}_{-0.03} \pm 0.01$~nb; and at 100 TeV energy, and find $3.25 \pm 0.20 \pm 0.04$ nb. We observe that at LHC energies the scale and pdf uncertainties are similar, but at higher collider energies the pdf uncertainties become smaller than the scale variation. 

\section{Differential distributions for $tW$ production}

We next consider differential distributions of the top quark and of the $W$ boson in $tW$ production. In particular, we calculate transverse-momentum and rapidity distributions. Differential distributions can be more sensitive to new physics than total cross sections, so it is important to consider the effect of soft-gluon corrections on these distributions.

\subsection{Top-quark differential distributions}

We first consider differential distributions in $p_T$ and rapidity of the top quark in $tW$ production.

\begin{figure}[htbp]
\begin{center}
\includegraphics[width=90mm]{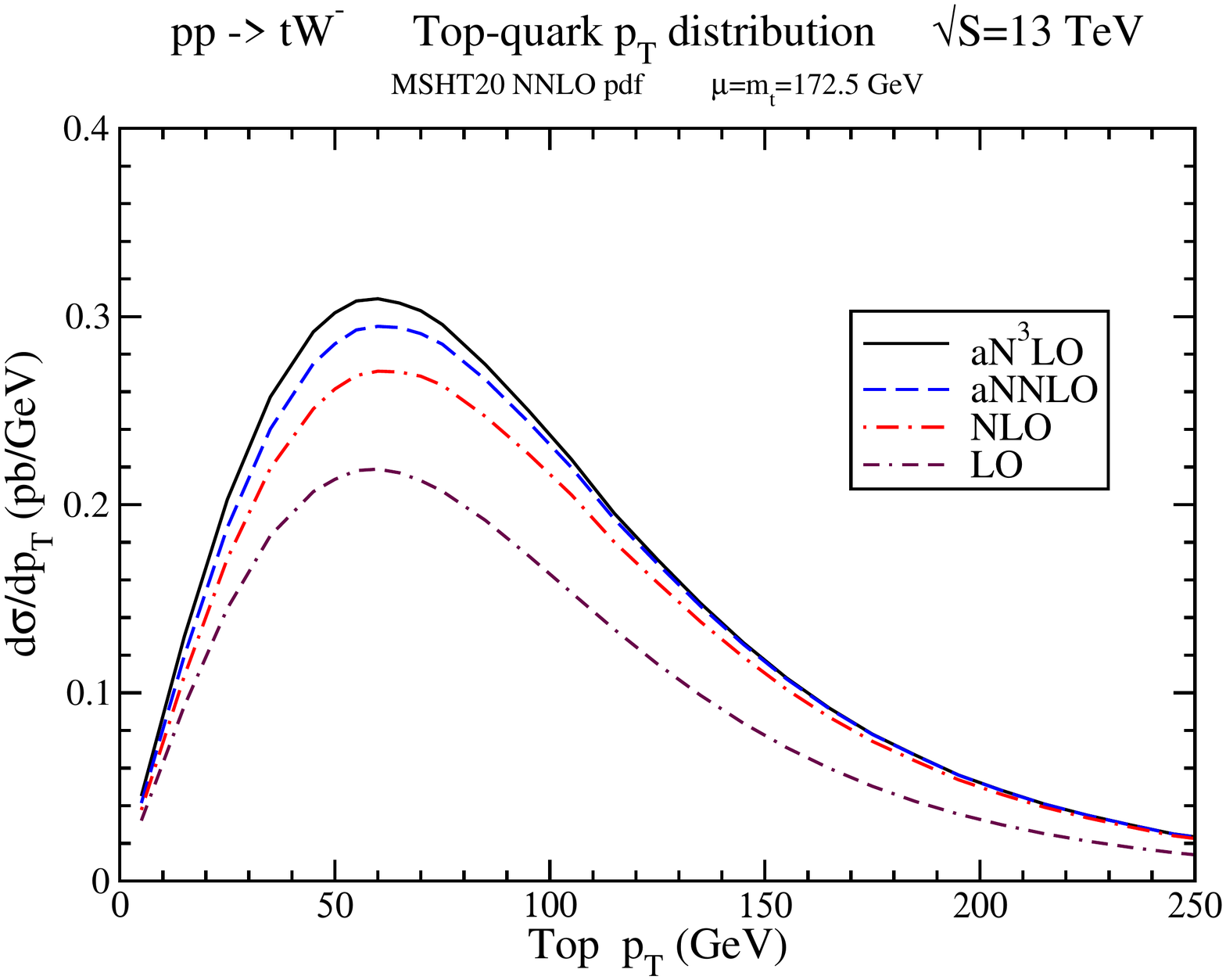}
\hspace{-7mm}
\includegraphics[width=90mm]{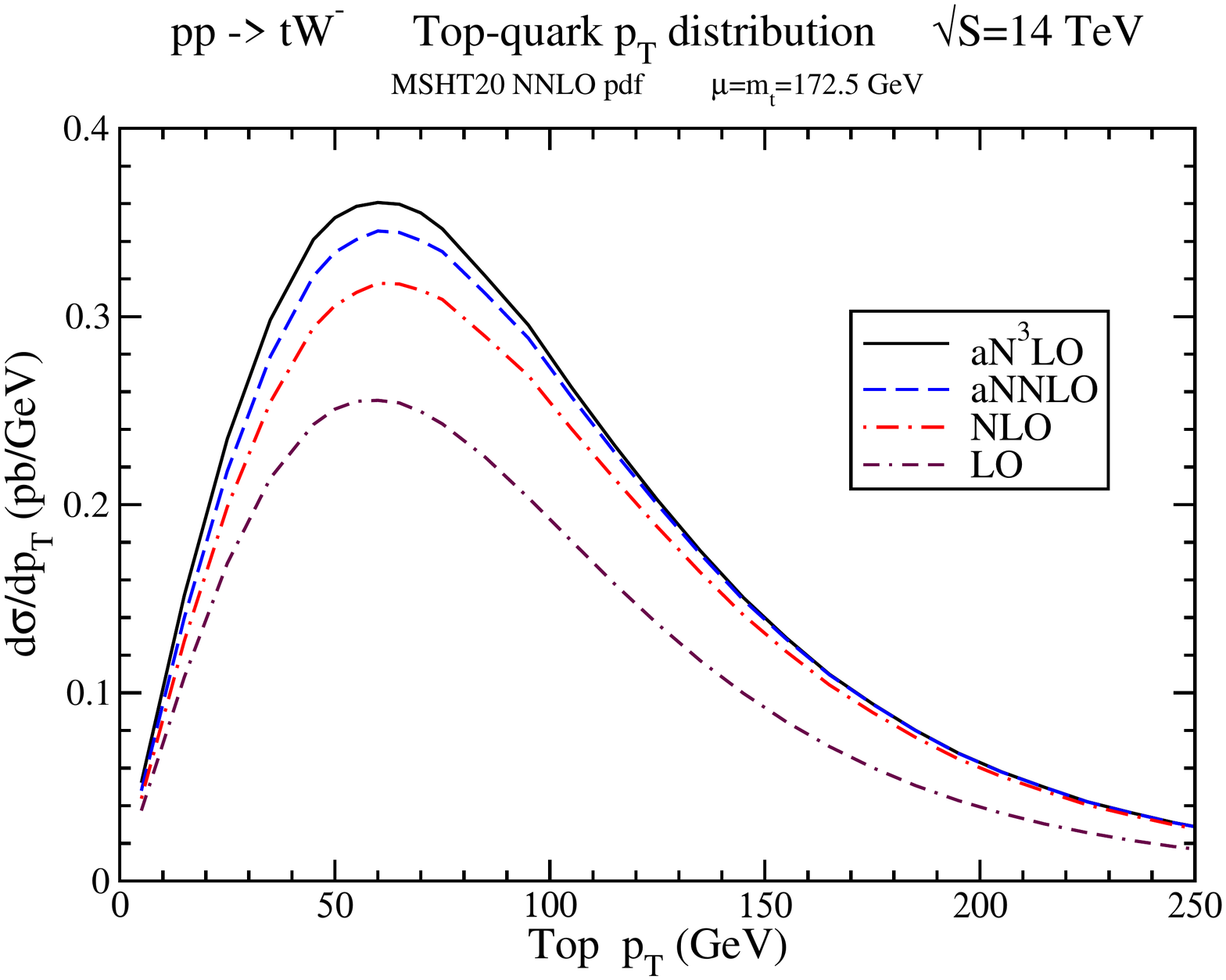}
\caption{The LO, NLO, aNNLO, and aN$^3$LO top-quark $p_T$ distributions in $tW^-$ production at 13 TeV (left) and 14 TeV (right) collider energies using MSHT20 NNLO pdf and $\mu=m_t=172.5$ GeV.}
\label{pttoptWplot}
\end{center}
\end{figure}

In Fig. \ref{pttoptWplot} we show the LO, NLO, aNNLO, and aN$^3$LO top-quark $p_T$ distributions, $d\sigma/dp_T$, at 13 TeV (left plot) and 14 TeV (right plot) collider energies. We have used MadGraph \cite{MadGraph} for the complete NLO $p_T$ distributions. The distributions for both energies peak at a $p_T$ value of around 60 GeV and they fall quickly at very high $p_T$ values. We observe that the NLO corrections to the $p_T$ distribution are quite large and that there are further significant enhancements from higher-order soft-gluon corrections. The aNNLO distribution is derived by adding the second-order soft-gluon corrections to the complete NLO result while the aN$^3$LO distribution is derived by also adding the third-order soft-gluon corrections. 

\begin{figure}[htbp]
\begin{center}
\includegraphics[width=90mm]{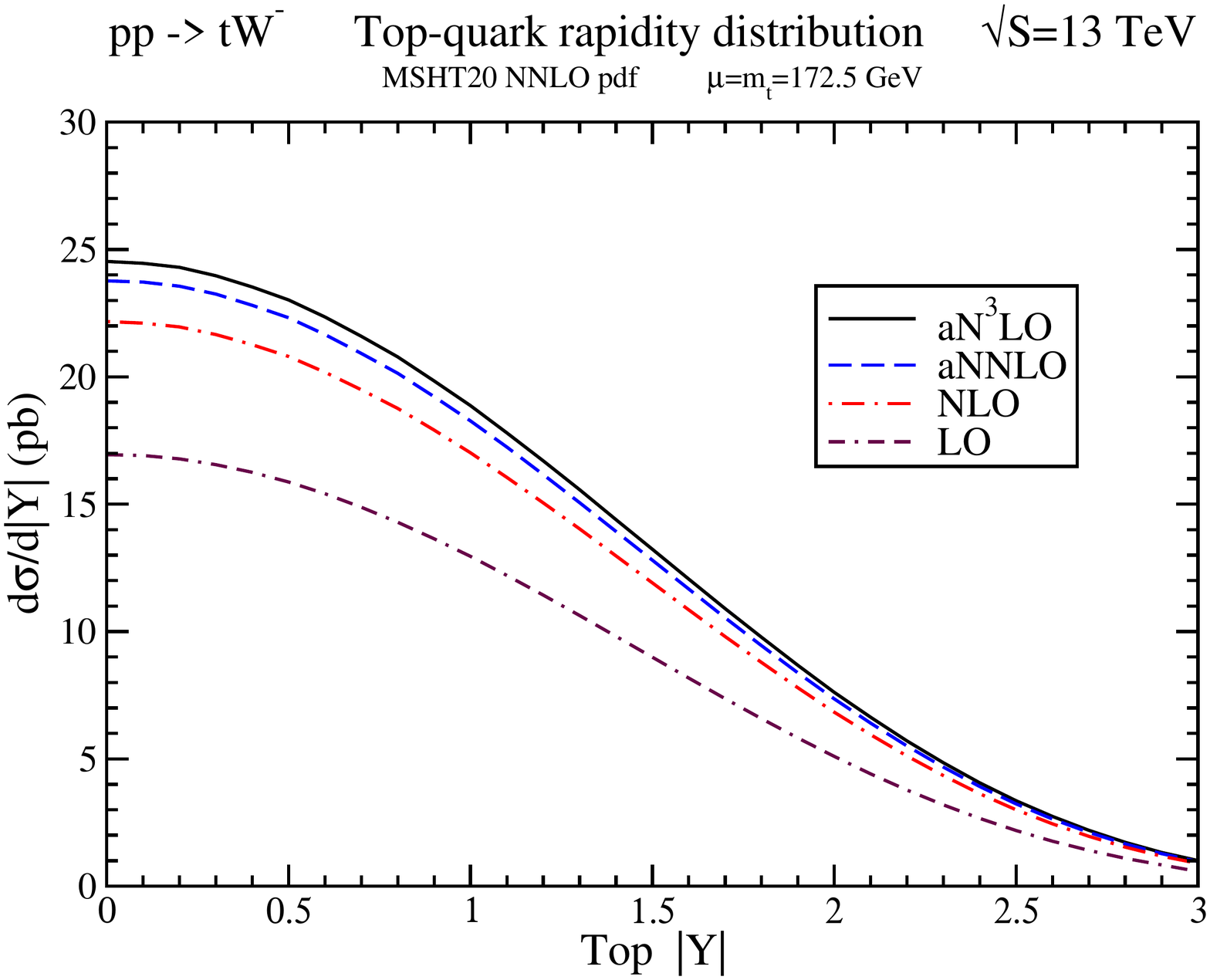}
\hspace{-7mm}
\includegraphics[width=90mm]{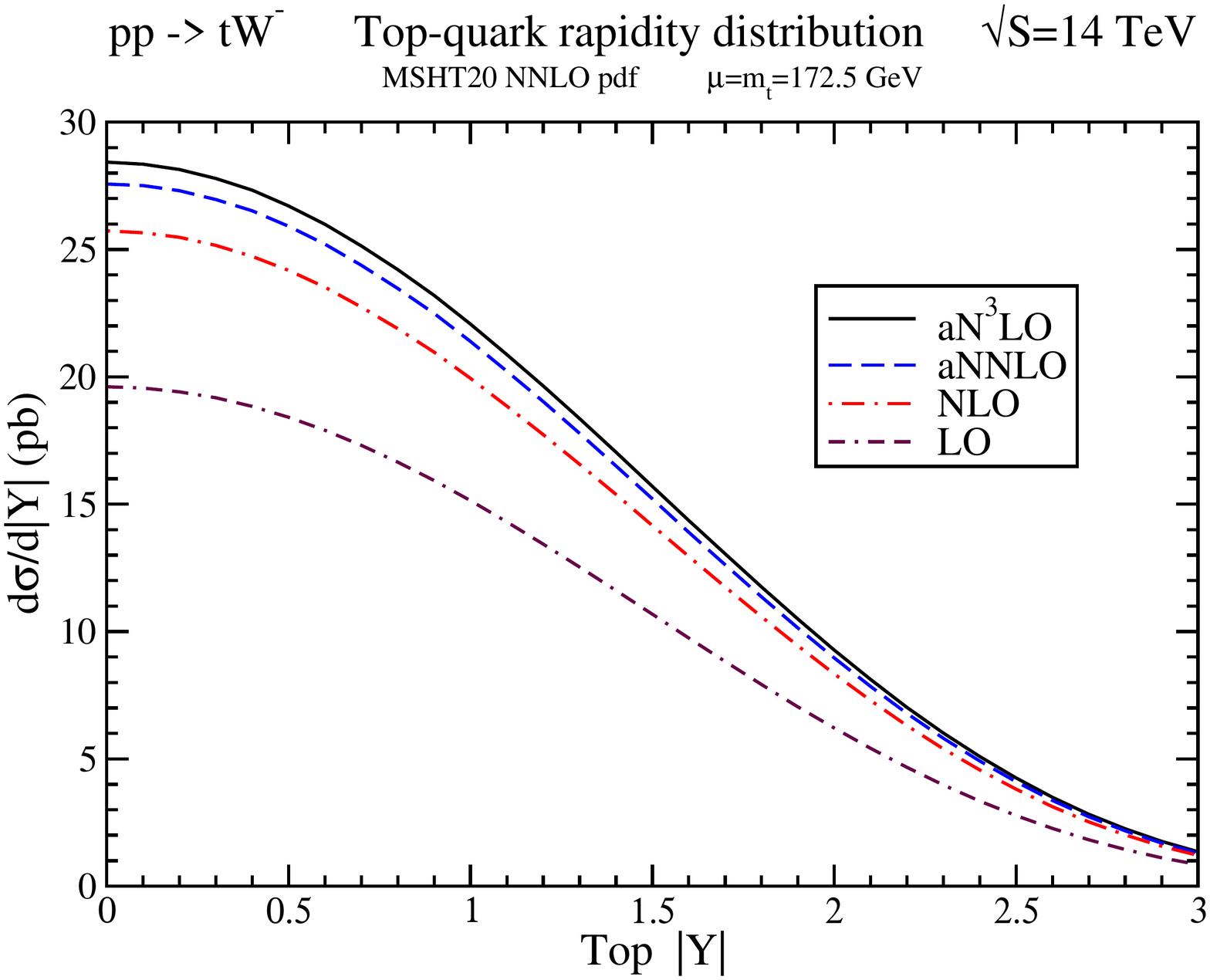}
\caption{The LO, NLO, aNNLO, and aN$^3$LO top-quark rapidity distributions in $tW^-$ production at 13 TeV (left) and 14 TeV (right) collider energies using MSHT20 NNLO pdf and $\mu=m_t=172.5$ GeV.}
\label{ytoptWplot}
\end{center}
\end{figure}

In Fig. \ref{ytoptWplot} we show the LO, NLO, aNNLO, and aN$^3$LO top-quark rapidity distributions, $d\sigma/d|Y|$, at 13 TeV (left plot) and 14 TeV (right plot) collider energies. We plot the distributions over a wide range up to a rapidity value of 3. We have used MadGraph \cite{MadGraph} for the complete NLO  rapidity distributions and added the soft-gluon corrections at second and third orders, as before, to obtain the aNNLO and aN$^3$LO distributions, respectively. Again, we observe large corrections at NLO and additional significant enhancements from the higher-order soft-gluon corrections at aNNLO and aN$^3$LO.

\begin{figure}[htbp]
\begin{center}
\includegraphics[width=90mm]{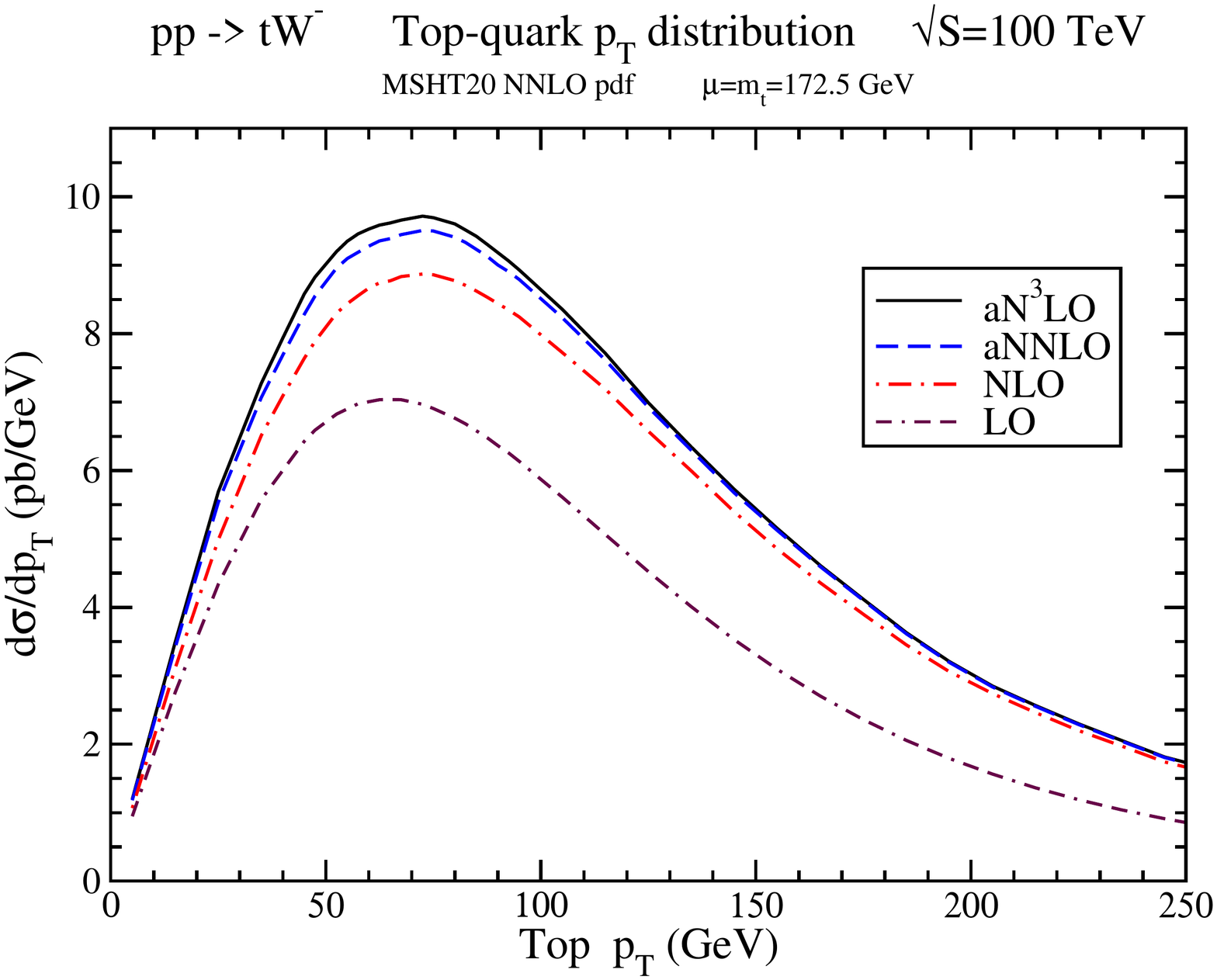}
\hspace{-7mm}
\includegraphics[width=90mm]{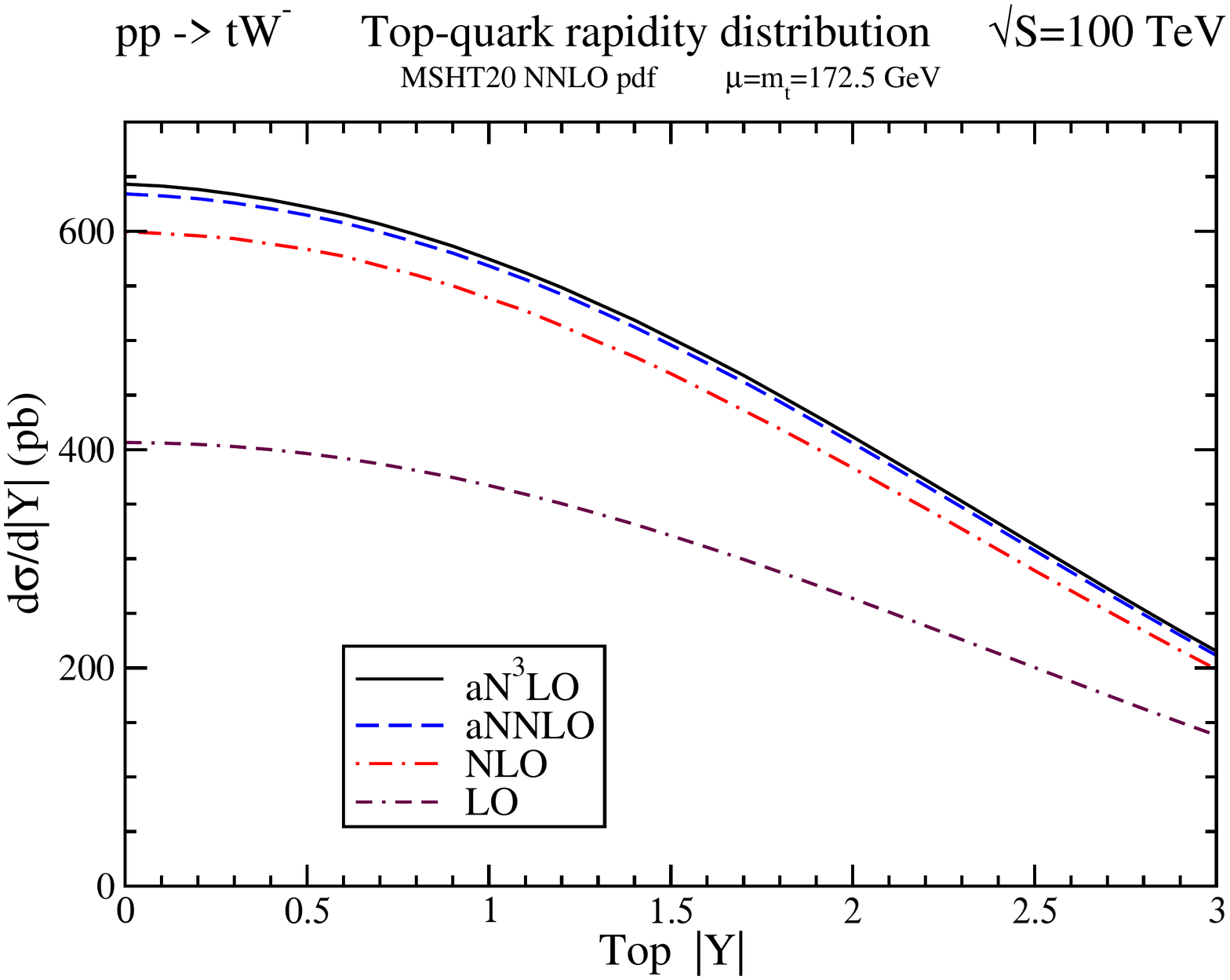}
\caption{The LO, NLO, aNNLO, and aN$^3$LO top-quark $p_T$ distributions (left) and rapidity distributions (right) in $tW^-$ production at 100 TeV collider energy using MSHT20 NNLO pdf and $\mu=m_t=172.5$ GeV.}
\label{ptytoptW100tevplot}
\end{center}
\end{figure}

In Fig. \ref{ptytoptW100tevplot} we show the LO, NLO, aNNLO, and aN$^3$LO top-quark $p_T$ distribution (left plot) and rapidity distribution (right plot) at 100 TeV energy. The $p_T$ distribution now peaks at a higher $p_T$ value of around 70 GeV. The rapidity distribution remains considerable even at a rapidity value of 3, in contrast to the distribution at LHC energies. Again, we observe large corrections at NLO and further nonnegligible contributions from the higher-order soft-gluon corrections in both differential distributions at 100 TeV collision energy.

\begin{figure}[htbp]
\begin{center}
\includegraphics[width=90mm]{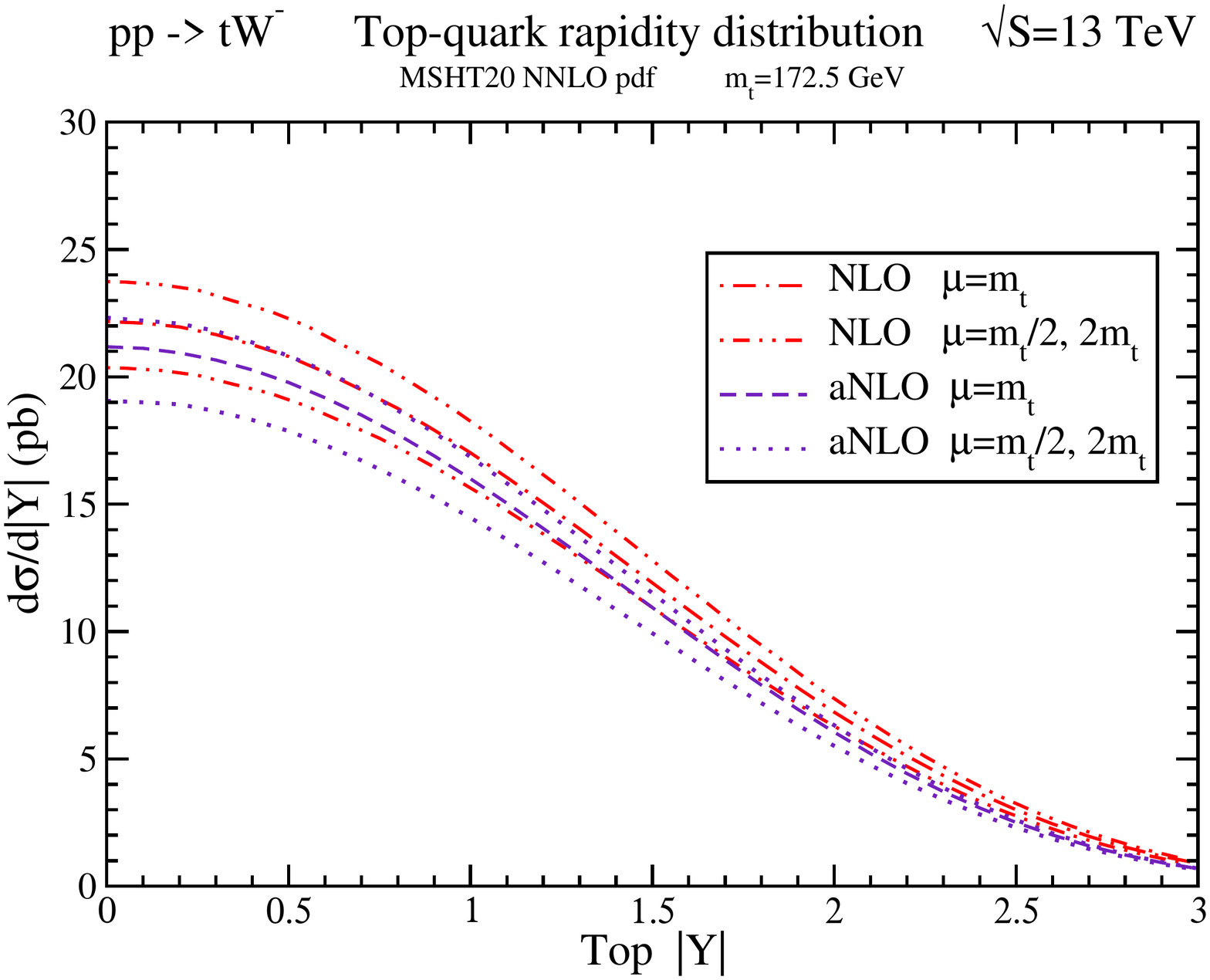}
\hspace{-7mm}
\includegraphics[width=90mm]{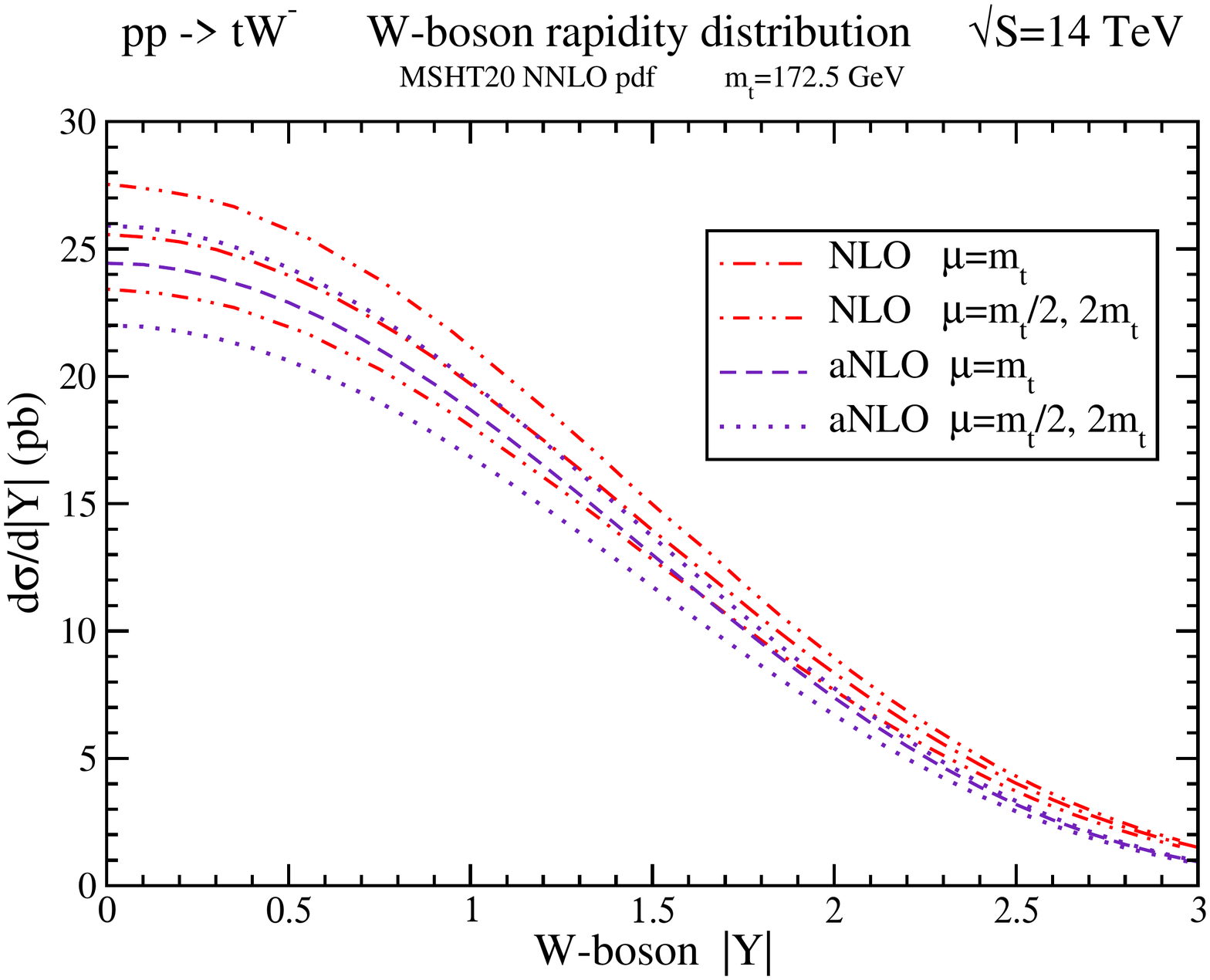}
\caption{The NLO and aNLO rapidity distributions in $tW^-$ production of (left) the top quark at 13 TeV collider energy and (right) the $W$ boson at 14 TeV collider energy using MSHT20 NNLO pdf with three scale choices: $\mu=m_t/2$, $m_t$, and $2m_t$.}
\label{ytWNLOplot}
\end{center}
\end{figure}

It should be noted that the soft-gluon approximation works very well not only for the total cross section, as already shown in the left plot of Fig. \ref{tWplot}, but also for the top-quark differential distributions in both magnitude and shape. As an example, on the left plot of Fig. \ref{ytWNLOplot} we compare the NLO and aNLO top-quark rapidity distributions at 13 TeV collider energy. We observe very good agreement between the NLO and aNLO results, for the central values as well as for the scale variation.

\subsection{$W$-boson differential distributions}

We next study the differential distributions in $p_T$ and rapidity of the $W$ boson in $tW$ production. 

We begin with noting the high quality of the soft-gluon approximation for the $W$-boson differential distributions, just as we did for the top-quark ones. As an example, on the right plot of Fig. \ref{ytWNLOplot} we compare the NLO and aNLO $W$-boson rapidity distributions at 14 TeV collider energy. Again, we observe very good agreement between NLO and aNLO, including scale variation.

\begin{figure}[htbp]
\begin{center}
\includegraphics[width=90mm]{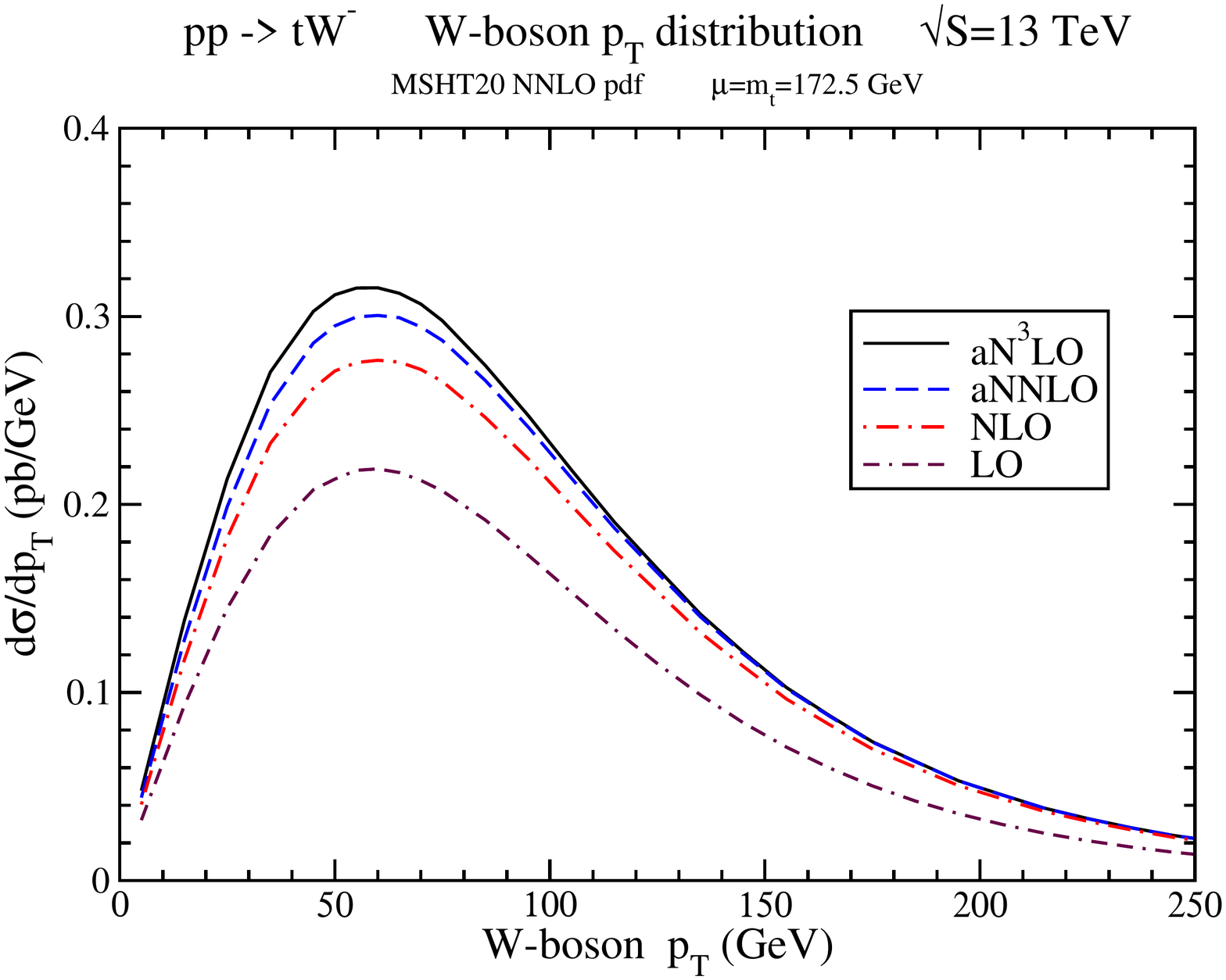}
\hspace{-7mm}
\includegraphics[width=90mm]{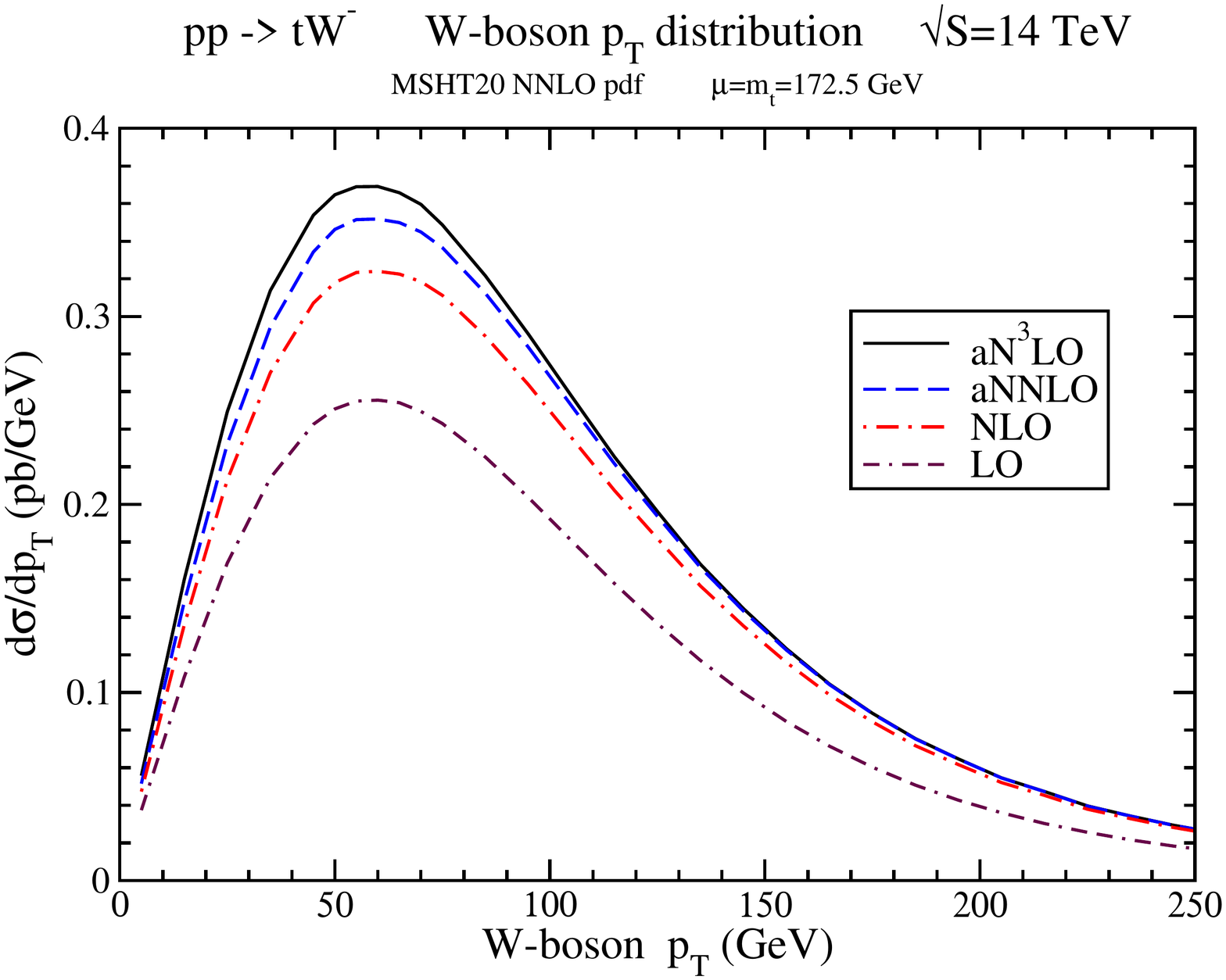}
\caption{The LO, NLO, aNNLO, and aN$^3$LO $W$-boson $p_T$ distributions in $tW^-$ production at 13 TeV (left) and 14 TeV (right) collider energies using MSHT20 NNLO pdf and $\mu=m_t=172.5$ GeV.}
\label{Wptplot}
\end{center}
\end{figure}

In Fig. \ref{Wptplot} we show the LO, NLO, aNNLO, and aN$^3$LO $W$-boson $p_T$ distributions, $d\sigma/dp_T$, at 13 TeV (left plot) and 14 TeV (right plot) collider energies. Again, we have used MadGraph \cite{MadGraph} for the complete NLO $p_T$ distributions. At LO the $W$-boson $p_T$ distribution is the same as that of the top quark (since the top quark and the $W$ boson are produced back-to-back) but the distributions begin to differ at NLO and higher. As before, we get the aNNLO distribution by adding the second-order soft-gluon corrections to the complete NLO result, and the aN$^3$LO distribution by further adding the third-order soft-gluon corrections. We note, again, that the higher-order corrections are significant.

\begin{figure}[htbp]
\begin{center}
\includegraphics[width=90mm]{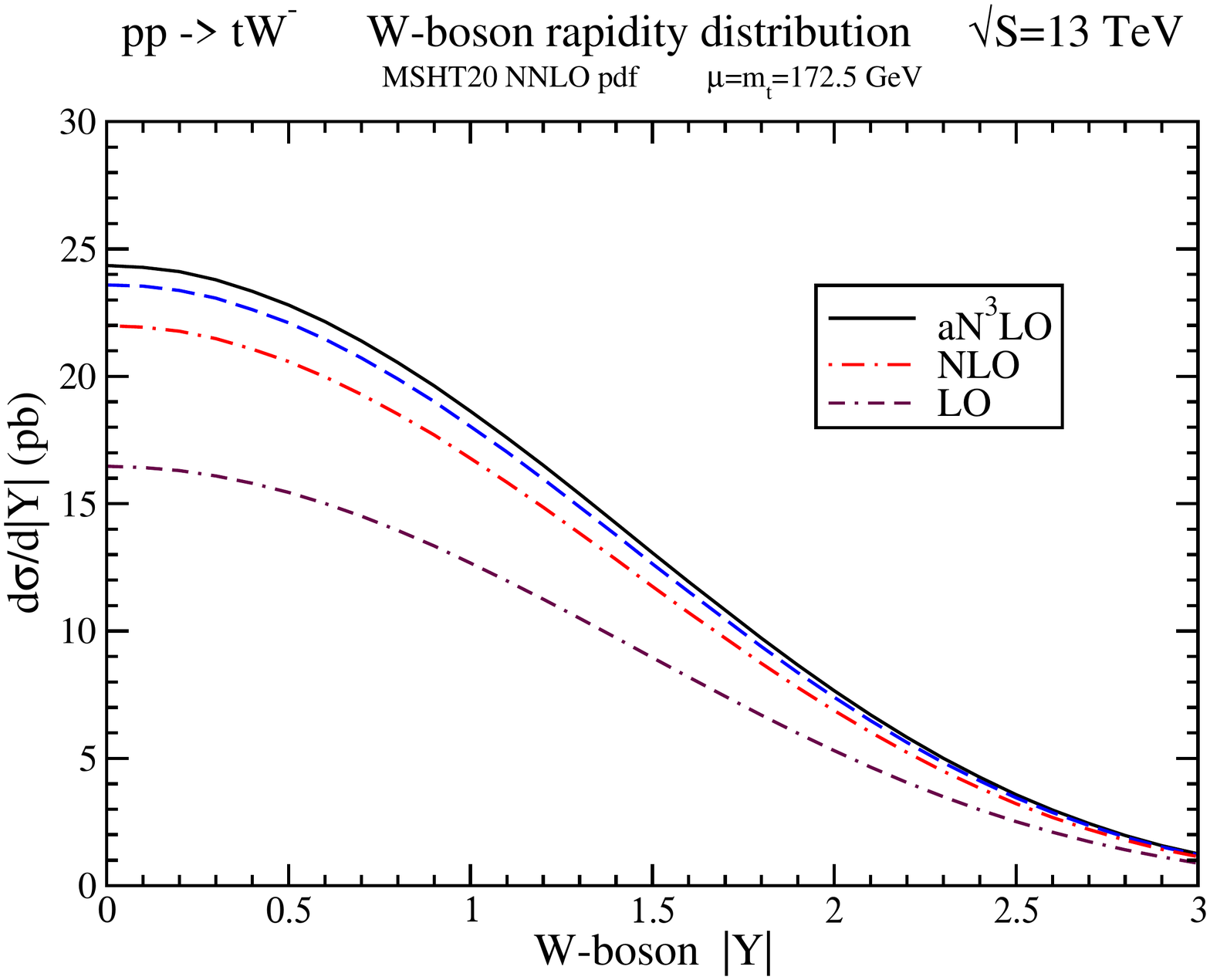}
\hspace{-7mm}
\includegraphics[width=90mm]{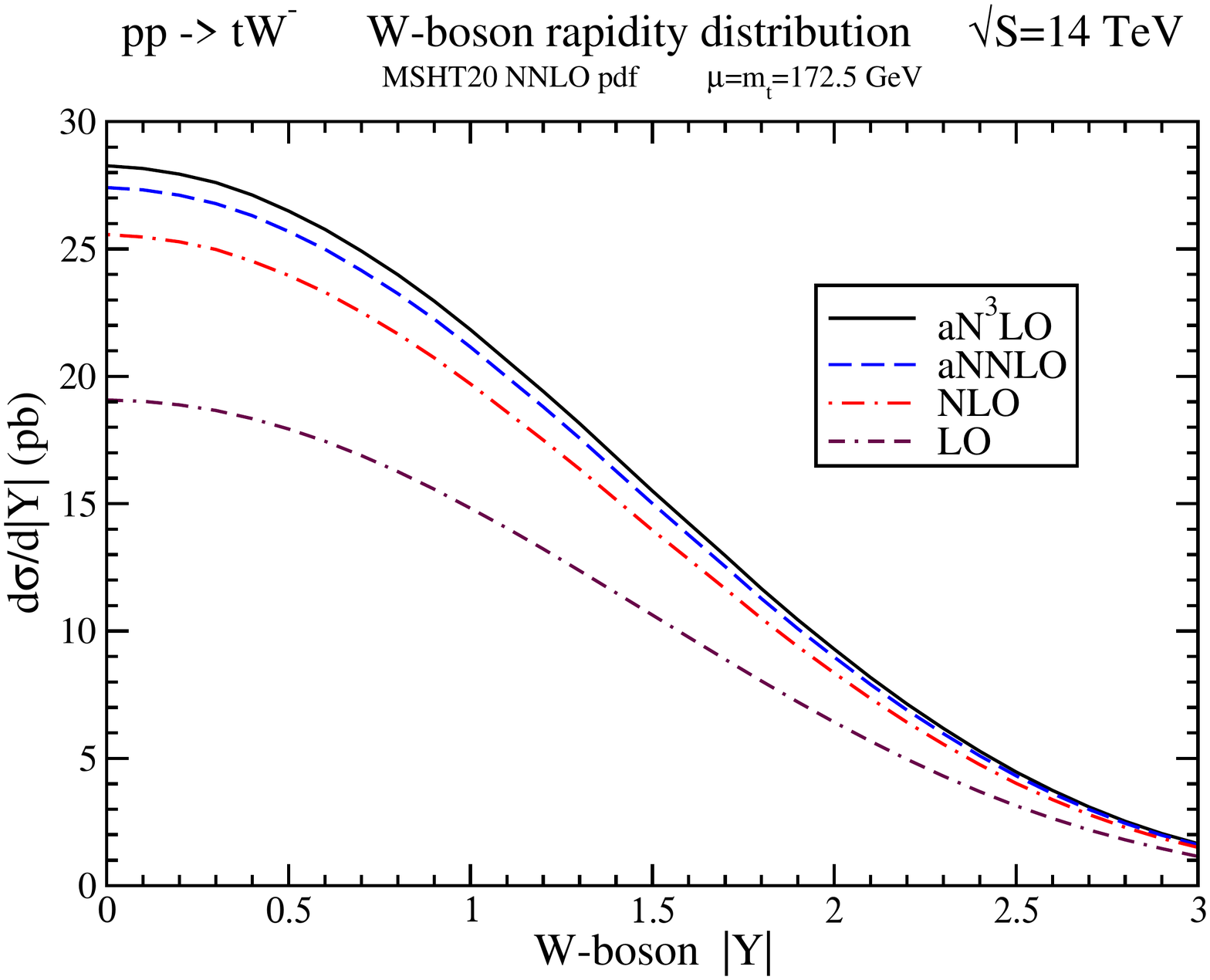}
\caption{The LO, NLO, aNNLO, and aN$^3$LO $W$-boson rapidity distributions in $tW^-$ production at 13 TeV (left) and 14 TeV (right) collider energies using MSHT20 NNLO pdf and $\mu=m_t=172.5$ GeV.}
\label{Wyplot}
\end{center}
\end{figure}

In Fig. \ref{Wyplot} we show the LO, NLO, aNNLO, and aN$^3$LO $W$-boson rapidity distributions, $d\sigma/d|Y|$, at 13 TeV (left plot) and 14 TeV (right plot) collider energies. Again, we plot the distributions over a wide range up to a rapidity value of 3. We observe significant contributions from the NLO corrections and from the additional soft-gluon corrections.

\begin{figure}[htbp]
\begin{center}
\includegraphics[width=90mm]{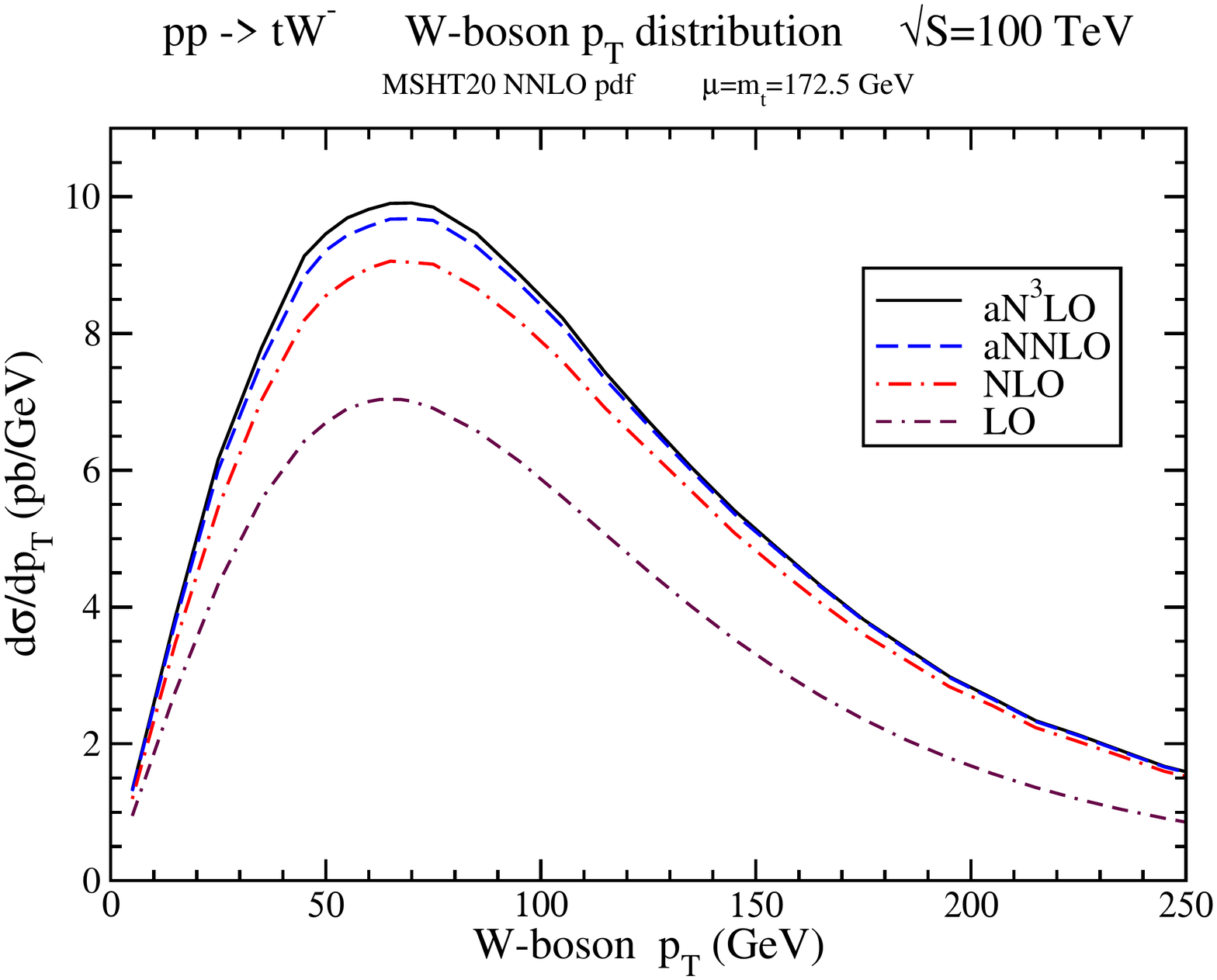}
\hspace{-7mm}
\includegraphics[width=90mm]{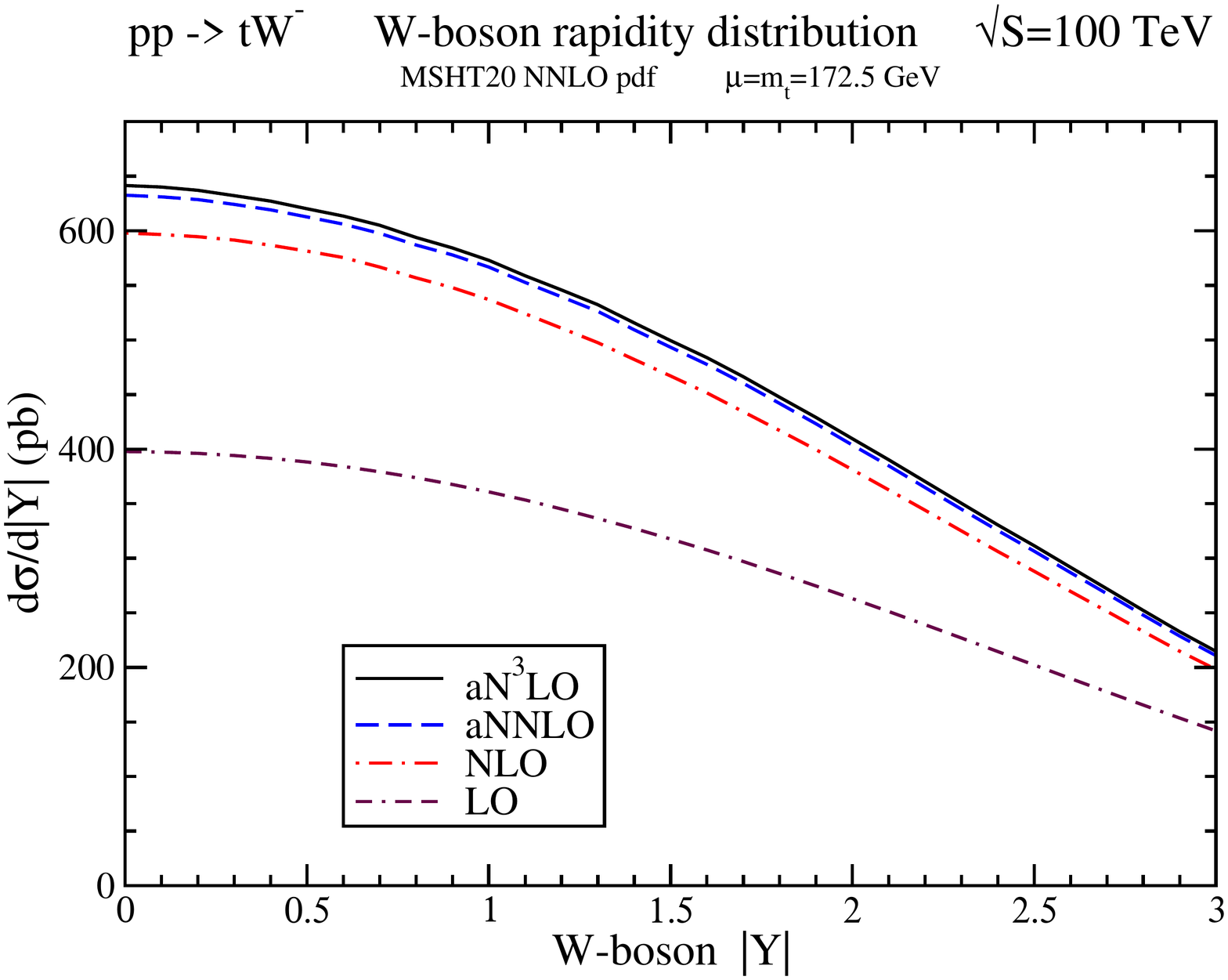}
\caption{The LO, NLO, aNNLO, and aN$^3$LO $W$-boson $p_T$ distributions (left) and rapidity distributions (right) in $tW^-$ production at 100 TeV collider energy using MSHT20 NNLO pdf and $\mu=m_t=172.5$ GeV.}
\label{Wpty100tevplot}
\end{center}
\end{figure}

Finally, in Fig. \ref{Wpty100tevplot} we show the LO, NLO, aNNLO, and aN$^3$LO $W$-boson $p_T$ distribution (left plot) and rapidity distribution (right plot) at 100 TeV energy, which again highlight the importance of the higher-order corrections. 

\mysection{Conclusions}

In this paper we have provided theoretical predictions for the associated production of a top quark with a $W$ boson in high-energy hadronic collisions. We have employed the soft-gluon resummation formalism to calculate higher-order corrections for this process. We have shown the high quality of the soft-gluon approximation for $tW$ production over a very wide range of energies at hadron colliders, including very high future energies up to 100 TeV. This indicates that the formalism is applicable to kinematical regions far from threshold, a conclusion with important implications for the study of $tW$ production and other top-quark processes.

We have presented results for the total cross sections for $tW$ production, including scale and pdf dependence, and have shown that they describe data from the LHC very well. We have also presented results for the top-quark and the $W$-boson tranverse-momentum and rapidity distributions. In all cases, the soft-gluon corrections at aNNLO and aN$^3$LO are significant and they decrease the uncertainty in the theoretical results.

\mysection*{Acknowledgements}
This material is based upon work supported by the National Science Foundation under Grant No. PHY 1820795.

\end{document}